%% file: main.tex
\documentclass[a4paper, 12pt]{article}
\usepackage{comment} 
\usepackage{graphicx}
\usepackage{caption}
\usepackage{color,soul}
\usepackage{fullpage} 
\usepackage{natbib}
\usepackage[T1]{fontenc}
\usepackage{upquote}
\usepackage[dvipsnames]{xcolor}
\setcitestyle{authoryear,open={(},close={)}}
\usepackage{hyperref}
\usepackage{amsmath,amsfonts,amssymb}
\usepackage{longtable}
\usepackage{color,soul}
\usepackage[toc,page]{appendix}
\usepackage{booktabs}
\usepackage{nicematrix}
\usepackage[section]{placeins}
\usepackage{multirow}
\usepackage{tikz}
\usetikzlibrary{fit}
\usepackage{colortbl}
\usepackage{import}
\usepackage{float}
\usepackage{setspace}
\DeclareRobustCommand{\hlcyan}[1]{{\sethlcolor{SkyBlue}\hl{#1}}}
\DeclareRobustCommand{\lime}[1]{{\sethlcolor{lime}\hl{#1}}}
\DeclareRobustCommand{\gold}[1]{{\sethlcolor{Goldenrod}\hl{#1}}}
\DeclareRobustCommand{\apricot}[1]{{\sethlcolor{Apricot}\hl{#1}}}
\DeclareRobustCommand{\orch}[1]{{\sethlcolor{Thistle}\hl{#1}}}

\begin{document}
\title{Estimating temporary emigration from capture-recapture data in the presence of latent identification}
\author{Katarina Skopalova$^1$ \and Jafet Belmont$^1$ \and Wei Zhang$^2$}
\date{$^1$School of Mathematics and Statistics, University of Glasgow, Glasgow, UK\\
$^2$School of Mathematics and Statistics, University of Sydney, Sydney, NSW, Australia}
\maketitle

\begin{abstract}
\noindent
Most capture-recapture models assume that individuals either do not emigrate or emigrate permanently from the sampling area during the sampling period. This assumption is violated when individuals temporarily leave the sampling area and return during later capture occasions, which can result in biased or less precise inferences under normal capture-recapture models. Existing temporary emigration models require that individuals are uniquely and correctly identified. To our knowledge, no studies to date have addressed temporary emigration in the presence of latent individual identification, which can arise in many scenarios such as misidentification, data integration, and batch marking. In this paper, we propose a new latent multinomial temporary emigration modelling framework for analysing capture-recapture data with latent identification. The framework is applicable to both closed- and open-population problems, accommodates data with or without individual identification, and flexibly incorporates different emigration processes, including the completely random and Markovian emigration. Through simulations, we demonstrate that model parameters can be reliably estimated in various emigration scenarios. We apply the proposed framework to a real dataset on golden mantella collected using batch marks under Pollock's robust design. The results show that accounting for temporary emigration provides a better fit to the data compared to the previous model without temporary emigration. 

\end{abstract}
\newpage

\input{intro}
\input{models_methods}

\input{simulation}

\input{mantella}

\input{discussion}
\input{acknowledgement}

%\clearpage
%=============================================
\bibliographystyle{acl}

\input{main.bbl}
%\bibliography{Cites.bib}

\clearpage
\doublespacing      
\subimport{supplementary_material/}{sm_main.tex}

\end{document}

%% file: intro.tex
\section{Introduction}
Capture–recapture (CR) studies are widely used in wildlife conservation and management to estimate demographic parameters such as abundance, survival, and recruitment. A common assumption underlying most CR models \citep[e.g.][]{otis1978statistical,cormack1964estimates,jolly1965explicit,seber1965note} is that any emigration (death or leaving) from the study area during sampling is permanent. In practice, this assumption is violated in many situations where individuals temporarily leave the study area \citep{kendall1997estimating}. For example, when sampling is conducted at breeding sites, non-breeding individuals may be absent and thus undetectable. Similarly, when sampling does not cover the entire population habitat, individuals may move out of the study area and then return, making them not always available for capture during the sampling period. When temporary emigration occurs but is not addressed properly in CR models, the resulting estimators of key model parameters could be biased or less precise \citep{kendall1997estimating,kendall1999robustness}.\\

Most temporary emigration models in the literature were developed for CR data collected under the robust design \citep{pollock1982capture}. In such studies, individuals are sampled across multiple primary periods, between which the population is assumed to be open to gains (i.e. birth and immigration) and losses (i.e. death and emigration). Each primary period comprises a number of secondary sampling occasions, during which the population is assumed to be closed. The robust design makes it possible to estimate temporary emigration from data over the secondary occasions, which provide additional information on capture probabilities that would otherwise be confounded with emigration in standard open-population CR models \citep{kendall2002estimating}. With the robust design CR data, \citet{Kendall1995} and \citet{kendall1997estimating} defined two temporary emigration models: (1) completely random emigration, where the probability that an individual stays out of the study area at time $t+1$ does not depend on its emigration status (i.e. in or out of the study area) at time $t$; (2) first-order Markovian emigration, where the emigration probability at time $t+1$ depends on whether the individual was a temporary emigrant at time $t$. \citet{schwarz1997estimating} relaxed the assumption of population closure within primary periods and developed an open robust design model. The model allows individuals to emigrate or immigrate between secondary occasions, but can only incorporate the completely random emigration. \citet{kendall2002estimating} later extended it to allow the Markovian emigration.\\

Without the robust design, it is not possible to estimate all model parameters in the presence of temporary emigration, unless certain constraints are imposed on some parameters \citep{fujiwara2002general}. For example, one can require some model parameters to be constant over time, use partially deterministic between-state transitions, or assume the same survival probability for different emigration stages. \cite{fujiwara2002general} presented an approach based on the rank of the Jacobian of the log-likelihood function for determining the estimability of model parameters. Using an analytic–numerical approach, \citet{kendall2002estimating} and \citet{schaub2004estimating} provided a more comprehensive assessment of parameter estimability for the model of \citet{fujiwara2002general} and more general variants. More recent approaches to modelling temporary emigration include mixture models, which assume that the population is composed of different groups with similar capture probabilities \citep{matechou2023capture} and models that use additional sources of information about species migratory patterns, such as telemetry data \citep{bird2014estimating}.\\

% Without the robust design, the estimation of all model parameters is not possible unless some constraints are imposed \citep{fujiwara2002general}. For example, one can require some model parameters to be constant over time, use partially deterministic between-state transitions, or assume the same survival probability for both the observable and unobservable states. \cite{fujiwara2002general} presented an approach based on the rank of the Jacobian of the log-likelihood function for determining the estimability of model parameters. Using an analytic-numerical approach, \cite{kendall2002estimating} and \cite{schaub2004estimating} investigated more thoroughly the parameter estimability issue for the model of \cite{fujiwara2002general} and more general variants. More recent approaches to modelling temporary emigration include mixture models, which assume that the population is composed of different groups with similar capture probabilities \citep{matechou2023capture} and models that use additional sources of information about species migratory patterns, such as telemetry data \citep{bird2014estimating}.\\

The temporary emigration models described above all require CR data with individual identification and do not readily extend to cases where individual identification is latent. Latent identification in CR studies refers to situations where individual identification is missing or uncertain. This can arise in many situations, for example, when individuals are misidentified \citep[e.g.][]{link2010uncovering}, different datasets are integrated into one analysis \citep[e.g.][]{bonner2013mark}, and batch marks are used for sampling \cite[e.g.][]{zhang2022latent}. Our work in this paper was motivated by the analysis of a batch-marking dataset on golden mantella (\textit{Mantella aurantiaca}) shown in \cite{zhang2022latent}. In their analysis, \citet{zhang2022latent} identified a near-seasonal pattern in the estimates of capture probabilities over the whole sampling period. The authors suggested that this might reflect individuals leaving and re-entering the study area, but they did not explore this further. Given that temporary emigration, if not accounted for, could bias key model parameters like survival, we aim to develop a temporary emigration model that can deal with latent identification. \\

In this paper, we present a new latent multinomial modelling framework for analysing CR data with temporary emigration, where individual identification may or may not be available. In the latent multinomial model (LMM), the true but latent demographic (recruitment and emigration) and capture processes are modelled using a multinomial distribution, and the observed data vector is connected to the latent vector through a deterministic linear transformation. The first LMM was proposed by \cite{link2010uncovering} for modelling misidentification in CR data, and was later developed for problems in various fields including ecology, epidemiology, and population genetics \citep[e.g.][]{bonner2013mark,mcclintock2013integrated,zhang2019fast,zhang2021maximum,zhang2022latent,yong2024}. In this paper, we demonstrate the proposed model framework on batch-marking data with temporary emigration in open populations. We also illustrate how the model framework can be applied to problems with other types of latent identification in CR studies. We check by simulations that model parameters can be reliably estimated for CR data either with or without individual identification. We then apply the proposed temporary emigration model to analyse the golden mantella data of \cite{zhang2022latent}.\\

This paper is organised as follows. In Section \ref{mod}, we describe the latent multinomial temporary emigration model, briefly introduce the model fitting approach and discuss the parameter identifiability issues. Section \ref{sims} presents simulation studies assessing the performance of the proposed modelling framework. Section \ref{app} shows the results of fitting the proposed temporary emigration model to the mantella data and provides a comparison to the model of \cite{zhang2022latent} without temporary migration. The paper concludes with a discussion and ideas for future research in Section \ref{dis}.

%% file: models_methods.tex
\section{Models and Methods}\label{mod}
\subsection{Setup}\label{setup}
The modelling framework presented in this paper is designed for data collected under the robust design, as is the case for the motivating mantella dataset. In Section \ref{modelvariants}, we outline how the framework could, in principle, be adapted to accommodate data from other study designs.\\

We consider a CR study with $K$ primary periods indexed as $k=1,\ldots,K$, where each primary period $k$ has $T_k$ secondary occasions indexed as $l=1,\ldots,T_k$. In total, there are $T = \sum^{K}_{k=1}T_{k}$ capture occasions. Similar to the POPAN model of \cite{schwarz1996general}, we define the following model parameters:
\begin{itemize}
    \item $N$ - superpopulation size, i.e. the number of all individuals that are ever available for capture during the study;
    \item $\gamma_{k}$ - the probability that a new individual enters the superpopulation between primary periods $k-1$ and $k$. Let $\boldsymbol{\gamma}=(\gamma_1, \ldots, \gamma_K)^{\intercal}$. Note that $\sum^{K}_{k=1}\gamma_{k}=1$ and thus there are $K-1$ free parameters. Following \cite{zhang2022latent}, we introduce $\gamma_1^{*},\ldots,\gamma_{K-1}^{*}$ such that $\gamma_1^{*}\!=\!\gamma_1$, $\gamma_2^{*}\!=\!\gamma_2/(1\!-\!\gamma_1)$, $\gamma_3^{*}\!=\!\gamma_3/(1\!-\!\gamma_1\!-\!\gamma_2)$, ..., $\gamma_{K-1}^{*}\!=\!\gamma_{K-1}/(1\!-\!\gamma_1\!-\!\ldots\!-\!\gamma_{K-2})$. These represent the conditional probabilities that an individual becomes first available for capture during primary period $k$, given that it was unavailable for capture previously; 
    \item $\phi_k$ - the probability that an individual survives from primary period $k$ to $k+1$. Let $\boldsymbol{\phi}=(\phi_1, \ldots, \phi_{K-1})^{\intercal}$;
    \item $p_{kl}$ - the probability that an individual is captured on secondary occasion $l$ of primary period $k$, given that the individual is alive and in the study area. Let $\boldsymbol{p}=(p_{11}, \ldots, p_{KT_K})^{\intercal}$ contain all these capture probabilities. 
\end{itemize}

Following \cite{kendall1997estimating}, we define the following parameters related to temporary emigration:
\begin{itemize}
\item $\alpha_k$ = \text{Pr(alive, not in population in period $k+1$ | alive, in population in period $k$)} for $k=1,\ldots,K-1$. $\boldsymbol{\alpha}=(\alpha_1, \ldots, \alpha_{K-1})^{\intercal}$;
\item $\beta_k$ = \text{Pr(alive, in population in period $k+1$ | alive, not in population in period $k$)} for $k=2,\ldots,K-1$. $\boldsymbol{\beta}=(\beta_2, \ldots, \beta_{K-1})^{\intercal}$;
\item $\alpha_k'$ = \text{Pr(alive, not in population in period $k+1$ | alive in period $k$)} for $k=1,\ldots,K-1$. $\boldsymbol{\alpha}'=(\alpha_1',\ldots,\alpha_{K-1}')^{\intercal}$.
\end{itemize}

The parameters $\boldsymbol{\alpha}$ and $\boldsymbol{\beta}$ define the Markovian emigration model, which assumes that the probability of becoming a temporary emigrant for an individual in one primary period depends on the individual's emigration status in the previous period. When $\alpha_k = 1-\beta_k$ for $k=2,\ldots,K-1$, then the Markovian model reduces to the completely random emigration model with parameters $\boldsymbol{\alpha}',$ where the probability of becoming a temporary emigrant does not depend on the previous emigration status. \\

We also make the following assumptions: (1) both temporary and permanent emigration (i.e. death) only occur between primary periods; (2) the population is closed within each primary period; (3) survival probability is the same for individuals in and outside of the study area; (4) individuals do not lose their tags during the study; and (5) individuals are independent from each other.

\subsection{Latent Process}
To describe the true but unobservable status of each individual in the superpopulation on each secondary sampling occasion, we define the following latent states: 
\begin{itemize}
    \item 0 - the individual has not joined the population yet;
    \item 1 - the individual is not captured;
    \item 2 - the individual is captured;
    \item 3 - the individual is temporarily unavailable for capture; and 
    \item 4 - the individual has died or left the population permanently. 
\end{itemize}
A sequence of these states over $T$ secondary occasions forms the latent capture history of an individual. These five latent states result in a total of $5^T$ possible latent histories. However, not all latent histories are valid, as the order of the states matters. A latent history is excluded if: (1) it is only composed of state 0, i.e. the individual has never joined the population; (2) state 0 is followed by a state other than 0 within a given primary period; (3) state 0 appears after any of the other four states; (4) the history starts with states 3 or 4; (5) state 3 is preceded by a state other than 3 in the given primary period (e.g. a primary period with states `133' would be invalid, as emigration has to happen before the start of the period); (6) state 3 is followed by any state other than 3 within a given primary period; (7) state 4 is followed by any state other than 4.\\

Let $J$ denote the total number of \emph{valid} latent histories, which we index by $j=1, \ldots, J$. Let $\pi_j=\pi_j(\boldsymbol{\theta})$ be the probability that an individual has latent history $j$, which depends on a parameter vector $\boldsymbol{\theta}$ consisting of $\boldsymbol{\gamma}, \boldsymbol{\phi}, \boldsymbol{p},$ and $(\boldsymbol{\alpha}, \boldsymbol{\beta})$ or $\boldsymbol{\boldsymbol{\alpha}'}$. Define $z_j$ to be the number of individuals with latent history $j$. Assuming independence between all the $N$ individuals yields that the latent vector of counts
\begin{equation}
\label{eqn:eqn1}
\boldsymbol{z} \sim \text{Multinomial}(N, \boldsymbol{\pi}),
\end{equation}
where $\boldsymbol{z}=(z_1, \ldots, z_J)^{\intercal}$ and $\boldsymbol{\pi}=(\pi_1, \ldots, \pi_J)^{\intercal}$.\\

To illustrate how to obtain the vector $\boldsymbol{\pi}$ of latent capture probabilities in terms of $\boldsymbol{\theta}$, we define three transition probability matrices. Transition probability matrix $\boldsymbol{\Gamma}_{kl}$ between secondary occasions $l$ and $l+1$ within primary period $k$ is given in (\ref{eqn:transmat-1}): 
\begin{equation}
\label{eqn:transmat-1}
\boldsymbol{\Gamma}_{kl} = 
\begin{NiceTabular}{c@{\;}ccccc}
& & \Block{1-4}{Occasion $l+1$} \\ 
  &   & 1   & 2   & 3  & 4\\
\Block{4-1}{\rotate Occasion $l$} 
  & 1  & $1-p_{k,l+1}$  & $p_{k,l+1}$  & 0 & 0\\
  & 2 & $1-p_{k,l+1}$ & $p_{k,l+1}$ & 0 & 0\\
  & 3 & 0 & 0 & 1 & 0\\
  & 4 & 0 & 0 & 0 & 1\\
\CodeAfter \SubMatrix[{3-3}{6-6}].
\end{NiceTabular}
\end{equation}
Transition probabilities from the last secondary occasion of primary period $k$ to the first secondary occasion of period $k+1$ are given in (\ref{eqn:transmat-2}) for the completely random emigration model:
\begin{equation}
\label{eqn:transmat-2}
\boldsymbol{\Gamma}_{k}^{R} = 
\begin{NiceTabular}{c@{\;}ccccc}
& & \Block{1-4}{Period $k+1$} \\ 
  &   & 1   & 2   & 3  & 4\\
\Block{4-1}{\rotate Period $k$} 
  & 1 & $\phi_k(1-\alpha'_k)(1-p_{k+1,1})$  & $\phi_k(1-\alpha'_k)p_{k+1,1}$  & $\phi_k\alpha'_k$ & $1-\phi_k$\\ 
       & 2 & $\phi_k(1-\alpha'_k)(1-p_{k+1,1})$ & $\phi_k(1-\alpha'_k)p_{k+1,1}$ & $\phi_k\alpha'_k$ & $1-\phi_k$\\ 
       & 3 & $\phi_k(1-\alpha'_k)(1-p_{k+1,1})$ & $\phi_k(1-\alpha'_k)p_{k+1,1}$ & $\phi_k\alpha'_k$ & $1-\phi_k$\\ 
       & 4 & 0 & 0 & 0 & 1\\ 
\CodeAfter \SubMatrix[{3-3}{6-6}]
\end{NiceTabular}
\end{equation}
and in (\ref{eqn:transmat-3}) for the Markovian model:
\begin{equation}
\label{eqn:transmat-3}
\boldsymbol{\Gamma}_{k}^{M} = 
\begin{NiceTabular}{c@{\;}ccccc}
& & \Block{1-4}{Period $k+1$} \\ 
  &   & 1   & 2   & 3  & 4\\
\Block{4-1}{\rotate Period $k$} 
 & 1 & $\phi_k(1-\alpha_k)(1-p_{k+1,1})$  & $\phi_k(1-\alpha_k)p_{k+1,1}$  & $\phi_k\alpha_k$ & $1-\phi_k$\\ 
        & 2 & $\phi_k(1-\alpha_k)(1-p_{k+1,1})$ & $\phi_k(1-\alpha_k)p_{k+1,1}$ & $\phi_k\alpha_k$ & $1-\phi_k$\\ 
        & 3 & $\phi_k\beta_k(1-p_{k+1,1})$ & $\phi_k\beta_k p_{k+1,1}$ & $\phi_k(1-\beta_k)$ & $1-\phi_k$\\ 
        & 4 & 0 & 0 & 0 & 1\\ 
\CodeAfter \SubMatrix[{3-3}{6-6}].
\end{NiceTabular}
\end{equation}

% Note that state 0 is not included in these transition probability matrices as the calculation of latent capture probabilities starts from the first non-zero state (i.e. 1 or 2)---that is, once the individual becomes available for capture for the first time.\\ %The probability of the first non-zero state will include the entry probability $\gamma_k$ depending on the primary period during which the individual becomes available for capture.\\

We now consider how to calculate the probability $\pi_j$ of latent history $j$. Suppose the individual becomes first available for capture in primary period $f_j$. The individual can only enter the population in states 1 or 2, resulting in a probability of $\sigma_j$ that takes the value $\gamma_{f_j}(1-p_{f_j,1})$ for state 1 and $\gamma_{f_j}p_{f_j,1}$ for state 2. The probability $\pi_j$ is then computed as the product of $\sigma_j$, the within-period transition probabilities across all secondary occasions, and the between-period transition probabilities over subsequent primary periods. That is, we have
\begin{equation}
\pi_j = \sigma_{j} \times \prod_{k=f_j}^{K}\prod_{l=1}^{T_k-1} \tau_{kl} \times \prod_{k=f_j}^{K-1} \tau_{k},
\end{equation}
where $\tau_{kl}$ denotes the transition probability between secondary occasions $l$ and $l+1$ of period $k$ and $\tau_{k}$ the transition probability between primary periods $k$ and $k+1$. Based on the states on occasions $l$ and $l+1$ of period $k$, $\tau_{kl}$ is the corresponding entry in transition matrix $\Gamma_{kl}$. Similarly, according to the states on the last occasion of period $k$ and the first occasion of period $k+1$, $\tau_{k}$ are taken from either $\Gamma_k^R$ or $\Gamma_k^M$, depending on the emigration model used. As an example, consider a study with $K=4$ primary periods each with $T_k=2$ secondary occasions, and a latent history \{00 21 33 12\}. The probability of this latent history would be $\gamma_{2}p_{21}(1-p_{22})\alpha_2\beta_3(1-p_{41})p_{42}$ under the Markovian emigration model and $\gamma_{2}p_{21}(1-p_{22})\alpha'_2(1-\alpha'_3)(1-p_{41})p_{42}$ under the completely random emigration model.
%We can now express the probability of the $j$-th latent history $\pi_j=\pi_j({\theta})$ for each of the two emigration models in terms of their respective model parameters, i.e. ${\theta}=({\gamma},{\phi}, {p}, {\alpha}, {\beta})$ and ${\theta}=({\gamma},{\phi}, {p}, {\alpha})$ for the Markovian and completely random emigration model, respectively. Each $\pi_j=\pi_j({\theta})$ can be obtained as a product of the probability at entry (either $\gamma_{k}(1\!-\!p_{kl})$ or $\gamma_{k} p_{kl}$, depending on whether the individual first joins the population in state 1 or 2), and the probabilities of moving across the latent states for the remaining capture occasions.\\

%Since the model is structured as a robust design, where the population is assumed to be open between primary periods and closed within them, we define two `transition` matrices that describe the probabilities of moving between two latent states; one for transitions within and another for transitions between the primary periods. Both models share the same within-period transition matrix, which is shown in Table \ref{tab:tbl1}. The between-period transition matrices for the completely random and the Markovian emigration model are shown in Table \ref{tab:tbl2} and \ref{tab:tbl3}, respectively. Note that the between-period transition matrix for the completely random emigration model is just a special case of the between-period transition matrix of the Markovian emigration model with $\beta=1\!-\!\alpha$.\\

\subsection{Observed Data with Individual Identification}
When individual identification is available, the observed data (i.e. the capture history) for each individual is a vector of length $T$, where each element takes a value of 1 if the individual was captured on the given capture occasion and 0 otherwise. For example, an observed history of 0110 for a 4-occasion study means that the individual was captured on occasions 2 and 3 but not on occasions 1 and 4. There are $2^T \!- \!1$ distinct observable capture histories, excluding the null history, which is not observed. The observed data $\boldsymbol{y}$ is therefore a set of $2^T \! - \! 1$ frequencies of all the observable capture histories.

\subsection{Observed Data for Batch Marking}
Following the structure of the mantella dataset analysed in this paper, we consider batch-marking experiments where batch tags used remain the same for all secondary occasions within a given primary period, but differ by period.\\

Define $m_{kl}$ as the number of individuals first captured and marked on secondary occasion $l$ of primary period $k$. Similarly, define $n_{ktl}$ as the number of individuals subsequently recaptured on secondary occasion $l$ of primary period $t$, which were first marked in primary period $k$ ($k \leq t$). The observed vector of counts can then be represented as $\boldsymbol{y} = (\boldsymbol{m}^\intercal, \boldsymbol{n}^\intercal)^\intercal$, with $\boldsymbol{m}=(m_{11},\ldots,m_{1T_{1}},\ldots,m_{K1},\ldots,m_{KT_{K}})^\intercal$ and $\boldsymbol{n}=(n_{112},\ldots,n_{11T_{1}}, \ldots, n_{KK2}, \ldots, n_{KKT_{K}})^\intercal$. 

\subsection{Latent Multinomial Emigration Model}
One advantage of modelling the latent process using the latent states (0--4) is that it has a linear relationship with the observed data. This means that the latent random vector $\boldsymbol{z}$ and observed data $\boldsymbol{y}$ can be connected deterministically. More specifically, when individual identification is available, latent state 2 results in a 1 in the observed history, and all other latent states are observed as 0. Mapping latent and observed states for all capture occasions converts a latent history to an observed one. For example, latent histories \{0022331244\} and \{0022111211\} both produce the same observed history \{0011000100\}. Following this, a link matrix $\boldsymbol{A} = (a_{ij})$ can be derived so that 
\begin{equation}
    \boldsymbol{y} = \boldsymbol{A}\boldsymbol{z},
\end{equation}
where $a_{ij}=1$ if the $j$-th latent history yields the $i$-th observed history and 0 otherwise. \\

Derivation of the link matrix $\boldsymbol{A}$ is similar for batch-marking experiments. The entry $a_{ij}=1$ if the $j$-th latent history contributes the $i$-th count in the observed vector $\boldsymbol{y}$ and 0 otherwise. Specifically, for any latent capture history we need to find the primary period and secondary occasion on which the individual was first captured and marked, and primary periods and secondary occasions on which the individual was recaptured subsequently. For example, latent history \{12~21~33~12\} from a study with $K=4$ and $T_k=2$ (for $k=1,\ldots,4$) contributes one to the counts $m_{12}, n_{121},$ and $n_{142}$, as the individual was first captured and marked on occasion 2 of period 1 and recaptured on occasion 1 of period 2 and occasion 2 of period 4.\\

Since $\boldsymbol{z}$ follows a multinomial distribution as given in Equation (\ref{eqn:eqn1}), this is a latent multinomial model \citep{link2010uncovering,zhang2022latent}. 

%The latent vector of counts and the observed data are connected via the relationship ${y}={A}{z}$, where $A$ is a non-invertible but known matrix called the link matrix. In other words, each observed history directly arises from a non-invertible transformation of the latent histories given by the link matrix. For instance, assuming the robust design with five primary periods and two secondary occasions each, the latent histories \{0022331244\} or \{0022111211\} would both produce the same observed history: \{0011000100\}.\\

%For batch-marking data, the link matrix is formed by stacking up two matrices whose number of columns equals $J$, i.e. the number of valid latent histories. The first one corresponds to first captures and has a value of one at the $[i,j]$-th entry if the first capture of the $j$-th latent history occurred on the $i$-th capture occasion. The second matrix corresponds to recaptures and is composed of $K$ blocks of length $T$, i.e. $K\times T$ rows. If a recapture occurs on the $t$-th capture occasion in a particular latent history, and the first capture took place during the $k$-th primary period with $k=1,\ldots,K$, the $i$-th element of the $k$-th block is assigned a value of 1. Some rows of the link matrix will contain all zeros and can be omitted, as they are not needed.\\

\subsection{Parameter Estimation}
Our aim is to estimate the model parameters $N$ and $\boldsymbol{\theta}$ from the observed data $\boldsymbol{y}$, given $\boldsymbol{y}=\boldsymbol{A}\boldsymbol{z}$, where $\boldsymbol{z} \sim$ Multinomial$(N, \boldsymbol{\pi}(\boldsymbol{\theta}))$. This is not trivial, as the likelihood function of the LMM is not efficiently computable. Several approaches have been proposed for fitting LMMs \citep[e.g.][]{link2010uncovering,bonner2013mark,zhang2019fast}. For the reasons of computational efficiency, we follow the maximum likelihood method of \cite{zhang2019fast}, which uses the saddlepoint approximation. We do not introduce the method in detail here, but we encourage interested readers to refer to \cite{zhang2019fast} and \cite{zhang2022latent} for the methodology and implementation.\\

In our simulation study, we sometimes encountered the issue of extrinsic identifiability \citep{viallefont1998parameter}, where some parameters are estimated at or near the boundary of their defined range, (i.e. 0 or 1). This leads to a non-invertible or near singular Hessian matrix, and thus the produced standard errors become inflated. In such cases, confidence intervals tend to span the entire parameter space (0,1). Following \cite{zhang2022latent}, we use a penalised maximum likelihood approach with a penalty of the form $\mathcal{P}=\sum_{\theta \in \Theta_p} \operatorname{logit}(\theta)^2 /\left(2 \sigma_p^2\right)$ for each probability parameter, where $\Theta_p$ is the set of all probability parameters in the model and $\sigma_p$ is a given penalty tuning hyperparameter. Computationally, this form of penalty is equivalent to imposing independent priors on the parameters such that $\operatorname{logit}(\theta) \sim N(0, \sigma_p^2)$. In our analyses, we set $\sigma_p=2.5$ which works well for the simulated and real data, but the choice is adjustable.

\subsection{Parameter Identifiability}
The proposed latent multinomial emigration model suffers from identifiability issues when both survival and migration parameters are time-dependent, regardless of whether individual identification is available in the observed data. Specifically, the final survival and emigration parameter $\phi_{K-1}$ and $\alpha'_{K-1}$ are confounded in the completely random emigration model. Similarly, $\alpha_{K-1}$, $\beta_{K-1}$ and $\phi_{K-1}$ are not separately identifiable in the Markovian emigration model. These identifiability issues are similar to the confounding of the final capture and survival encountered in the traditional Cormack-Jolly-Seber model.\\

% When transitioning from the penultimate to the final primary period (i.e. from period $K-1$ to period $K$) in the completely random emigration model, the transition probabilities from state 1/2/3 to states 1 and 2 all involve the term $\phi_{K-1}(1-\alpha_{K-1}')$, while transitioning from state 1/2/3 to states 3 and 4 involves $\phi_K\alpha_{K-1}'$ and $1-\phi_{K-1}$ (see Equation \ref{eqn:transmat-2}). Note that states 3 and 4 cannot be distinguished from observed data because both yield a non-observation. This is not a problem for primary periods before the last one, because state 2 appearing in later periods after state 3 results in an observation which helps to distinguish state 3 from state 4. If an individual enters state 4, it will stay in state 4 for the remaining occasions. Thus, the transitions from state 1/2/3 to states 3 and 4 can be combined with a probability of $1-\phi_{K-1}(1-\alpha_{K-1}')$. As a result, $\phi_{K-1}$ and $\alpha_{K-1}'$ always appear together in the product $\phi_{K-1}(1-\alpha_{K-1}')$ and thus cannot be estimated separately.\\

When transitioning from the penultimate to the final primary period (i.e. from period $K-1$ to period $K$) in the completely random emigration model, the transition probabilities from state 1/2/3 to states 3 and 4 involve $\phi_K\alpha_{K-1}'$ and $1-\phi_{K-1}$ (see Equation \ref{eqn:transmat-2}). In this case, however, states 3 and 4 cannot be distinguished from each other based on the observed data, as both yield a non-observation. Note that this is not an issue for primary periods before the last one, as state 2 appearing after state 3 in later periods results in an observation, and if an individual enters state 4, it will remain in state 4 for the remaining periods. Thus, the transitions from state 1/2/3 to states 3 and 4 can be combined with a probability of $1-\phi_{K-1}(1-\alpha_{K-1}')$. As a result, $\phi_{K-1}$ and $\alpha_{K-1}'$ always appear together in the product $\phi_{K-1}(1-\alpha_{K-1}')$ and therefore cannot be estimated separately.\\

For the Markovian emigration model, the transition probabilities from state 1/2 to states 3 and 4 are given by $\phi_{K-1}\alpha_{K-1}$ and $1-\phi_{K-1}$ (see Equation \ref{eqn:transmat-3}), summing up to $1-\phi_{K-1}(1-\alpha_{K-1})$ due to the lack of information in data to distinguish states 3 and 4 in the last primary period. In addition, the transition probabilities from state 3 to states 3 and 4 sum up to $1-\phi_{K-1}\beta_{K-1}$. It follows that parameters $\phi_{K-1}, \alpha_{K-1}, \beta_{K-1}$ are not fully identifiable, and instead the products $\phi_{K-1}(1-\alpha_{K-1})$ and $\phi_{K-1}\beta_{K-1}$ are identifiable.\\

% original text: For the Markovian emigration model, as seen from Table 3 the product $\phi_{K-1}(1-\alpha_{K-1})$ appears in the transition probabilities from state 1/2 to states 1 and 2, while the transition probabilities from state 1/2 to states 3 and 4 are $\phi_{K-1}\alpha_{K-1}$ and $1-\phi_{K-1}$, summing up to $1-\phi_{K-1}(1-\alpha_{K-1})$ due to the lack of information in data to distinguish states 3 and 4 in the last primary period. In addition, the transition probabilities from state 3 to states 1 and 2 involve $\phi_{K-1}\beta_{K-1}$, while the transition probabilities from state 3 to states 3 and 4 sum up to $1-\phi_{K-1}\beta_{K-1}$. It follows that parameters $\phi_{K-1}, \alpha_{K-1}, \beta_{K-1}$ are not fully identifiable, and instead the products $\phi_{K-1}(1-\alpha_{K-1})$ and $\phi_{K-1}\beta_{K-1}$ are identifiable.\\

Note that these identifiability issues only affect fully time-dependent models. The models could be made fully identifiable with some constraints on the time-dependent parameters, for example, any one of $\phi_k, \alpha_k, \alpha_{k}'$ and $\beta_k$ is constant over different primary periods.

\subsection{Model Variants}\label{modelvariants}
The model described above can be readily adapted to accommodate different data scenarios for both closed and open populations. Modifications could also be made to allow other types of latent identification, such as misidentification and data integration. Below we will briefly describe the modifications needed to adjust the model to each of these scenarios.\\

Closed-population models assume that the population is closed to gains (i.e. births and immigration) and losses (i.e. deaths or emigration). To adapt the proposed framework to this setting, we specify: (1) $\gamma_1 = 1$ and $\gamma_2=\cdots=\gamma_K=0$, i.e. all individuals are available for capture from the beginning of the study; and (2) survival probabilities $\phi_k=1$ for $k=1,\ldots,K-1$. As noted by \cite{kendall1999robustness}, abundance estimators based on common closed-population models are biased in the case of Markovian emigration and have reduced precision in the case of random emigration. \cite{kendall1999robustness} considered CR data with individual identification, while our work makes it possible to address the issue for data with latent identification.\\

In an open-population scenario with no robust design, the model should allow immigration and emigration to occur between any two secondary occasions. This could be accommodated by specifying an entry probability $\gamma_t$ for each capture occasion with $\sum_{t=1}^{T}\gamma_t=1$, and $\phi_t$ as the survival probability from occasion $t$ to occasion $t+1$ for $t=1,\ldots,T-1$. Similar to the multi-state model of \cite{fujiwara2002general}, our model will have similar identifiability issues and the strategies outlined in \cite{kendall2002estimating} and \cite{schaub2004estimating} can help to address these. \\

One can also consider other types of latent identification using the proposed framework, such as misidentification \citep{link2010uncovering} and data integration \citep{bonner2013mark}. More specifically, we can replace state 2  in the current framework with two separate states corresponding to `captured and correctly identified' and `captured but misidentified`. Following \cite{link2010uncovering}, one needs to define a parameter for misidentification rate which will be used to calculate the probabilities of latent capture histories together with other model parameters. Similarly, for data integration as considered in \cite{bonner2013mark}, latent state 2 in our model should be replaced with multiple states covering all possible capture events of different sampling methods, i.e. captured by one method or both. In both cases, the link matrix $\boldsymbol{A}$ will also need to be adjusted according to the mapping between the observed and latent histories.

%% file: simulation.tex
\section{Simulation Study}\label{sims}
Simulation studies were carried out to evaluate the proposed modelling framework under two data scenarios: batch-marking (BM) and individual identification (ID). Five temporary emigration (TE) models were evaluated, including the completely random emigration model, $\alpha_t'$, where the subscript $t$ indicates time-varying emigration probabilities, and four Markovian emigration models incorporating either time-varying or constant (denoted by subscript $c$) emigration probabilities: (1) $\alpha_c\beta_c$, (2) $\alpha_c\beta_t$, (3) $\alpha_t\beta_c$, and (4) $\alpha_t\beta_t$. Other model parameters in $\boldsymbol{\theta}$ were set to be time-dependent. In the BM scenario, we also compared the TE models to the non-temporary-emigration (NoTE) model of \citet{zhang2022latent} to investigate the effect on parameter estimation if existing temporary emigration is not properly modelled. \\

Here we only present the results of one simulation setting with batch marking data under the fully time-dependent Markovian TE model $\alpha_t\beta_t$. More results for other TE models and simulation settings can be found in the supplementary material. The super-population size was set to $N=5000$, which is similar to that for the mantella data analysis shown below in this paper. The remaining model parameters were set to: $p_{kl}=0.4$, $\phi_k=0.9$, $\gamma_k=1/6$, $\alpha_k=0.2$, and $\beta_k=0.7$. We generated 100 datasets under the robust design with $K=6$ primary periods and $T_k=2$ secondary occasions within each primary period. \\ %Each model was fitted using both the original and penalised likelihoods, which we denote by subscripts $u$ and $p$, respectively.\\ 

\begin{figure}
    \centering
    \includegraphics[width=1\linewidth]{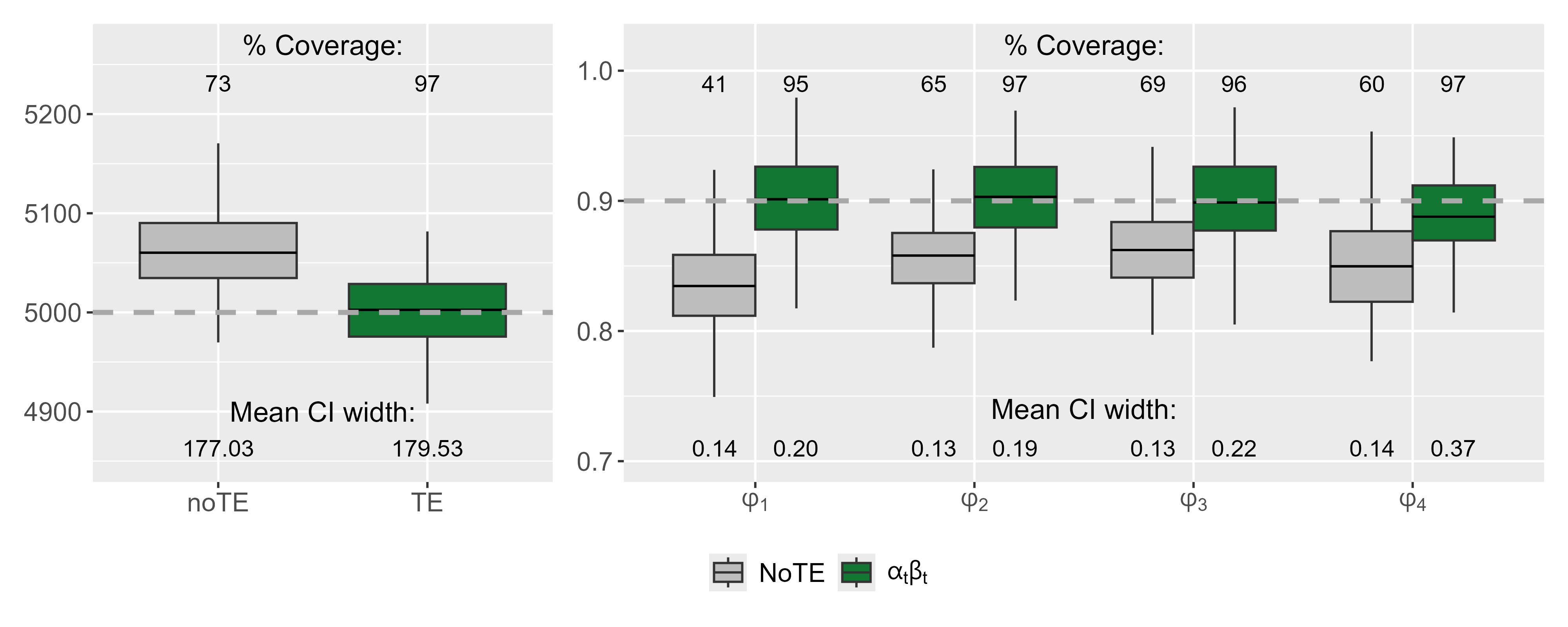}
    \caption{Boxplots of maximum likelihood estimates of super-population size $N$ (left) and survival probabilities $\boldsymbol{\phi}$ (right) obtained by fitting the NoTE model and the Markovian TE model $\alpha_t\beta_t$ to 100 datasets simulated under the Markovian TE model. The horizontal line inside each box represents the mean estimate across 100 simulations, while the gray dashed line indicates the true parameter value. 95\% CI coverage rates and mean CI widths are shown above and below each boxplot, respectively.}
    \label{fig:n_comp}
\end{figure}

Figure \ref{fig:n_comp} displays the results of the super-population size $N$ (left panel) and survival probabilities (right panel), obtained by fitting the NoTE model and the true Markovian TE model $\alpha_t\beta_t$ to simulated data from the Markovian TE model. In the left panel of Figure \ref{fig:n_comp}, the mean (5002.58) of the estimates of $N$ is very close to the true value (5000) when the TE model $\alpha_t\beta_t$ is fitted to the data, with a 95\% confidence interval (CI) coverage rate of 0.97. Fitting the NoTE model to the data still gives a roughly unbiased estimator of $N$ with a mean estimate of 5060.24; however, the 95\% CI coverage 0.73 is much lower than the nominal value. The right panel of Figure \ref{fig:n_comp} shows the comparison between survival estimates from the NoTE and TE models. Note that the final survival parameter $\phi_5$ cannot be estimated separately due to non-identifiability, and therefore its results are not shown. The estimates of all survival probabilities from the TE model are centred around the true value with relatively narrow 95\% CIs and roughly nominal 95\% CI coverages. However, when temporary emigration is not accounted for under the NoTE model, survival probabilities tend to be noticeably underestimated, as temporary emigration cannot be distinguished from mortality. The corresponding 95\% CI coverages are all far below the nominal value. This indicates that when temporary emigration is not properly modelled in CR data with latent identification, inference results could be misleading. This coincides with the conclusion of \cite{kendall1997estimating} for CR data with individual identification.\\

%The simulation study results are presented below. We focus on comparisons between the NoTE model and two emigration models: the fully time-dependent completely random emigration model ($\alpha_t'$) and the first-order Markovian emigration model ($\alpha_t\beta_t$) on batch-mark data. Capture and entry probabilities are not reported, as the estimators were consistently unbiased with near-nominal coverage and relatively narrow 95\% confidence intervals. Figure \ref{fig:n_comp} presents a comparison of the mean estimates, 95\% confidence interval coverage, and mean 95\% confidence interval widths between the NoTE model and the emigration models. \\

We also investigated whether the emigration parameters can be estimated reliably from the TE model. Table \ref{tab:emig_res} presents the mean emigration parameter estimates along with 95\% CI coverages and mean CI widths. It can be seen that emigration probabilities $\boldsymbol\alpha$ are estimated with roughly no bias, nearly nominal CI coverage rates, and good precision that is reflected by the relatively narrow mean CI widths. The precision of the final emigration parameter $\alpha_4$ is slightly worse than others because available data for estimating the parameter become less as primary period number increases (i.e. fewer sampling occasions left). Slightly more negative bias is observed for the immigration probabilities $\boldsymbol{\beta}$, where $\beta_2$ has the highest percentage bias of about $-8.6\%$. As reflected by the large mean CI widths for the $\boldsymbol{\beta}$ estimates, the data do not contain enough information to estimate $\boldsymbol{\beta}$ precisely. As shown by \cite{kendall1997estimating}, estimation of $\boldsymbol{\beta}$ is also associated with poor precision for data with individual identification. It should be noted that in this simulation setup, we had $\alpha_k = 0.2$, indicating a relatively low probability for an individual to become emigrant between any two primary periods. Moreover, $\beta_k = 0.7$ indicates a high probability for emigrants to return to the population. It follows that at any time point the proportion of emigrants is relatively low, leading to the low precision in the estimation of $\boldsymbol{\beta}$. Larger values in $\boldsymbol{\alpha}$ or lower values in $\boldsymbol{\beta}$ yield a larger proportion of emigrants during sampling, and therefore more data will be available for a more precise estimation of $\boldsymbol{\beta}$. These findings are consistent with those of \citet{kendall1997estimating}. In addition, the precision of $\boldsymbol{\beta}$ estimation can be improved by imposing additional constraints on the emigration parameters, such as assuming constant $\boldsymbol{\alpha}, \boldsymbol{\beta}$, or both. This is not surprising as the model is simplified.\\

% \begin{table}[!ht]
% \small
%     \centering
%     \captionsetup{font=small}  
%     \caption{Emigration parameter estimates, 95\% confidence interval coverage, and 95\% confidence interval widths produced by fitting the Markovian emigration model ($\alpha_t\beta_t$), with $N=5,000$.}
%     \resizebox{0.5\textwidth}{!}{
%     \begin{tabular}{llrrrrrr}
%     \toprule
%        Parameter  & True Value & Mean & CIC\% & CIW\\
%        \midrule
%      $\alpha_1$ & 0.20 & 0.21 & 95 & 0.26 \\
%     $\alpha_2$ & 0.20 & 0.21 & 92 & 0.25 \\
%         $\alpha_3$ & 0.20 & 0.19 & 99 & 0.26 \\
%         $\alpha_4$ & 0.20 & 0.18 & 100 & 0.36 \\
%         $\beta_2$ & 0.70 & 0.64 & 99 & 0.74 \\
%         $\beta_3$ & 0.70 & 0.66 & 99 & 0.67 \\
%         $\beta_4$ & 0.70 & 0.68 & 99 & 0.70 \\
%         $\phi_5(1-\alpha_5)$ & 0.72 & 0.73 & 97 & 0.18 \\
%         $\phi_5\beta_5$ & 0.63 & 0.67 & 100 & 0.86 \\
%     \bottomrule
%     \end{tabular}}
%     \caption*{\scriptsize The simulation study involved generating 100 batch-mark datasets under the robust design with $K=6$, and $T_k=2$ for $k=1,\ldots,6$, and $N=5,000$. The `Mean' column contains the average of the parameter estimate across the simulations, `CIC\%' refers to the 95\% confidence interval coverage, and `CIW' denotes the average 95\% confidence interval width.} 
%     \label{tab:emig_res}
% \end{table}

\begin{table}
\small
    \centering
    \captionsetup{font=small}  
    \caption{Mean estimates of emigration parameters, 95\% confidence interval coverage rates (CIC\%), and mean 95\% confidence interval widths (CIW) produced by the Markovian TE model. The results are based on 100 batch-marking datasets generated under the robust design with $K=6$ and $T_k=2$ for $k=1,\ldots,6$. Other model parameter values for simulation are $N=5000, p_{kl}=0.4, \phi_k=0.9$ and $\gamma_k=1/6$.}
    \resizebox{0.55\textwidth}{!}{
    \begin{tabular}{lcccc}
    \toprule
       Parameter  & True Value & Mean & CIC\% & CIW\\
       \midrule
     $\alpha_1$ & 0.20 & 0.21 & 95 & 0.26 \\
     $\alpha_2$ & 0.20 & 0.21 & 92 & 0.25 \\
     $\alpha_3$ & 0.20 & 0.19 & 99 & 0.26 \\
     $\alpha_4$ & 0.20 & 0.18 & 100 & 0.36 \\
     $\beta_2$ & 0.70 & 0.64 & 99 & 0.74 \\
     $\beta_3$ & 0.70 & 0.66 & 99 & 0.67 \\
     $\beta_4$ & 0.70 & 0.68 & 99 & 0.70 \\
     $\phi_5(1-\alpha_5)$ & 0.72 & 0.73 & 97 & 0.18 \\
     $\phi_5\beta_5$ & 0.63 & 0.67 & 100 & 0.86 \\
    \bottomrule
    \end{tabular}}
    \label{tab:emig_res}
\end{table}

We focus on the proposed TE models for batch-marking data here. Simulation results for data with individual identification as shown in the Supplementary Material reveal that model parameters can also be reliably estimated under the proposed TE models. As expected, estimation is more precise when individual identification is available, since individual-level information is lost when batch-marking data are used. For survival and emigration parameters, ID and BM data yield similar estimates with roughly no bias for earlier primary periods. In later periods, although both scenarios show some bias, ID data generally produce estimates that are closer to the true values and more precise. For return probabilities $\boldsymbol{\beta}$, ID estimates tend to be more accurate than BM ones across all primary periods. Further details are provided in Section A.2 of the Supplementary Material.

%% file: mantella.tex
\section{Application}\label{app}
\subsection{Mantella data}
As described initially in \cite{zhang2022latent}, the mantella data were collected using Visible Implanted Elastomers, a type of batch mark, under the robust design over six primary periods from 2014 to 2016. The first three periods consisted of three secondary occasions each, and the remaining three each had four secondary occasions. The total numbers of individuals first marked during each of the six primary periods are $\boldsymbol{m}=(1090, 295,115,686,403,141),$
and the numbers of marked individuals recaptured over different primary periods are 

\begin{equation}
\label{eqn:recaptures}
\boldsymbol{\boldsymbol{n}} = 
\begin{NiceTabular}{c@{\;}ccccccc}
& & \Block{1-6}{Recapture period} \\ 
  &   & 1   & 2   & 3  & 4 & 5 & 6\\
\Block{6-1}{\rotate Marking period} 
  & 1  & 219  & 55  & 17 & 255 & 90 & 15\\
  & 2 & & 43 & 42 & 41 & 62 & 37\\
  & 3 & & & 35 & 7 & 2 & 0\\
  & 4 & & & & 174 & 81 & 30\\
  & 5 & & & & & 107 & 13\\
  & 6 & & & & & & 1\\
\CodeAfter \SubMatrix[{3-3}{8-8}].
\end{NiceTabular}
\end{equation}

The $k$-th row of matrix $\boldsymbol{n}$ gives the numbers of individuals that were marked in period $k$ and recaptured in the same period or afterwards. Given that there are three or four secondary occasions within each primary period, individuals marked on one secondary occasion of a primary period can be recaptured on later secondary occasions in the same period.\\

\subsection{Results}

We fitted five different TE models to the mantella data: the completely random emigration model $\alpha_t'$ and four Markovian emigration models: $\alpha_c\beta_c$, $\alpha_c\beta_t$, $\alpha_t\beta_c$, and $\alpha_t\beta_t$. Survival, entry, and capture probabilities were set to be time-dependent in all models. For comparison, we also present the results of the NoTE model of \cite{zhang2022latent} under the same settings. The models were compared using the Akaike information criterion (AIC) as used in \citet{zhang2022latent}.\\

AIC values along with the corresponding super-population size estimates from the fitted TE and NoTE models are listed in Table \ref{table:aics}. All the fitted TE models have a much lower AIC value than the NoTE model, which indicates that accounting for temporary emigration provides an obviously improved fit to the data. Among the TE models, AIC favours the fully time-dependent Markovian model $\alpha_t\beta_t$; however, the difference of AIC values between this model and model $\alpha_c\beta_t$ is very minor (1.54). Additional steps, such as model averaging, could be taken to combine the results from these two models; however this falls beyond the scope of this paper. From Table \ref{table:aics}, the point estimates of super-population size tend to be lower in the TE models compared to the NoTE model. However, there is considerable overlap in the 95\% confidence intervals for $\alpha_t\beta_t$, $\alpha'_t$, and the NoTE model, which suggests that the differences among these models are not substantial. In comparison, $\alpha_c\beta_t$ produces a much lower point estimate than the NoTE model with limited overlap between them.\\

\begin{table}[!ht]
    \caption{AIC values and super-population size estimates with corresponding 95\% confidence intervals for all models fitted to the mantella data.}
    \label{table:aics}
    \centering
    \resizebox{1\textwidth}{!}{\begin{tabular}{lcccc} \toprule
        {Model} & {Parametrisation} & {Number of parameters} & {AIC} & {$\hat{N}\text{(95\% CI)}$}\\ \midrule
         {Markovian} & {$\alpha_t\beta_t$}   & {40}  & {710.21} & {5232 (4728, 5863)}\\
         {} & {$\alpha_c\beta_t$}   & {37}  & {711.75} & {5026 (4726, 5371)}\\
         {} & {$\alpha_t\beta_c$}   & {38}  & {835.79} & {4990 (4713, 5306)}\\
         {} & {$\alpha_c\beta_c$}& {34}  & {836.05} & {5038 (4775, 5336)}\\\bottomrule
         {Completely random} & {$\alpha_t'$}   & {36}  & {759.63} & {5291 (4748, 5979)}\\\bottomrule
         {NoTE} & {} & {31} & {1029.34} & {5467 (5024, 5995)} \\
         \bottomrule
    \end{tabular}}
\end{table}

Table \ref{tab:alphabeta} shows the estimated emigration and return probabilities from the top two Markovian TE models $\alpha_t\beta_t$ and $\alpha_c\beta_t$. There are no obvious seasonal or other cyclical patterns in the estimated emigration probabilities from the $\alpha_t\beta_t$ model. The estimates of emigration probabilities from both models are generally very high (minimum 0.70) across the whole sampling period, indicating a high level of dynamics in the mantella population during the sampling period. For both models, the estimated probability of emigrants returning back to the population is very low (0.01) between primary periods 2 and 3 and becomes higher for later periods. Compared to the emigration parameters, the return probabilities are much lower, which means that many emigrants stayed out of the sampling area for more than one period. Considering the relatively narrow widths of the 95\% confidence intervals, most emigration and return probabilities are estimated with good precision in both models, except for $\alpha_4$ and $\beta_4$ in the $\alpha_t\beta_t$ model and $\beta_5$ in the $\alpha_c\beta_t$ model. This is expected, as for later primary periods, there are less data available to distinguish temporary emigration from mortality.\\

\begin{table}
    \centering
    %\captionsetup{font=small} 
    \caption{Emigration and return probability estimates (with 95\% confidence intervals) for the best two Markovian TE models $\alpha_t\beta_t$ and $\alpha_c\beta_t$ with the lowest AIC values for the mantella dataset.}
        \begin{tabular}{lll}
        \toprule
            Parameter & $\alpha_t\beta_t$ & $\alpha_c\beta_t$ \\
        \midrule
            $\alpha_1$  & 0.85 (0.79, 0.90)  & 0.82 (0.78, 0.86)\\
            $\alpha_2$  & 0.80 (0.69, 0.88)  & 0.82 (0.78, 0.86)\\
            $\alpha_3$  & 0.90 (0.78, 0.96)  & 0.82 (0.78, 0.86)\\
            $\alpha_4$  & 0.70 (0.41, 0.89)  & 0.82 (0.78, 0.86)\\
            $\alpha_5$  & -  & 0.82 (0.78, 0.86) \\
            \midrule
            $\beta_2$   & 0.01 (0.00, 0.05)   & 0.01 (0.00, 0.05) \\
            $\beta_3$   & 0.49 (0.37, 0.61) & 0.51 (0.38, 0.64)\\
            $\beta_4$   & 0.31 (0.11, 0.63) & 0.24 (0.14, 0.39)\\
            $\beta_5$   & -   & 0.34 (0.15, 0.59) \\
        \bottomrule
        \end{tabular}
    %\caption*{\scriptsize Notes: The parameter subscript $k$ for some parameter $\alpha_k$ represents emigrating after the $k$-th primary period, i.e. between $k$ and $k\!+1$. Similarly, $\beta_k$ represents immigrating after the $k$-th primary period, i.e. between $k$ and $k\!+1$. The entries corresponding to $\alpha_5$ and $\beta_5$ in the $\alpha_t\beta_t$ column are missing due to confounding issues.}
    \label{tab:alphabeta}
\end{table}

\begin{figure}
    \centering
    \includegraphics[width=1\linewidth]{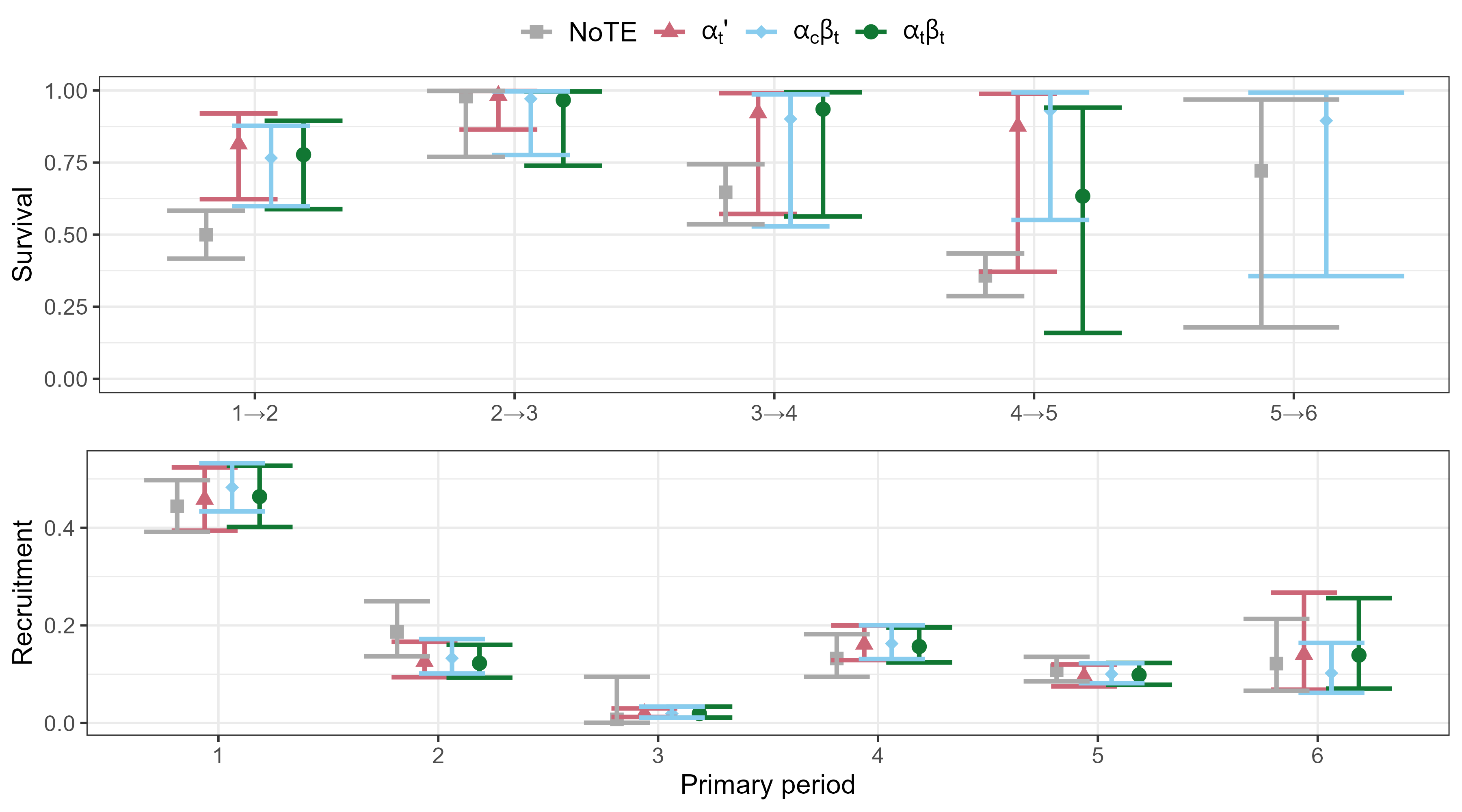}
    \caption{Estimation results of survival and recruitment parameters for the mantella data from the NoTE model and TE models $\alpha_t', \alpha_c\beta_t,$ and $\alpha_t\beta_t$.}
    \label{fig:phigamma}
\end{figure}

Estimation results of survival and recruitment parameters are shown in Figure \ref{fig:phigamma}. Point estimates and 95\% confidence intervals of recruitment parameters $\boldsymbol{\gamma}$ are similar between the NoTE and TE models, except for periods 2 and 3 for which the NoTE model yields higher recruitment probability estimates and wider confidence intervals than the TE models. Survival probabilities from TE models are generally higher than those of the NoTE model, which can be explained by the fact that temporarily emigrated individuals are no longer counted as permanent emigrants in TE models. Not surprisingly, there is more uncertainty in the survival estimates as time increases. This is expected given that available data become less in later periods to estimate survival parameters, while emigration and return probabilities need to be estimated simultaneously. Overall, the TE model $\alpha_c\beta_t$ seems to provide more stable estimates of survival parameters than the model $\alpha_t\beta_t$, although the latter has a slightly lower AIC value.\\

\begin{figure}
    \centering
    \includegraphics[width=1\linewidth]{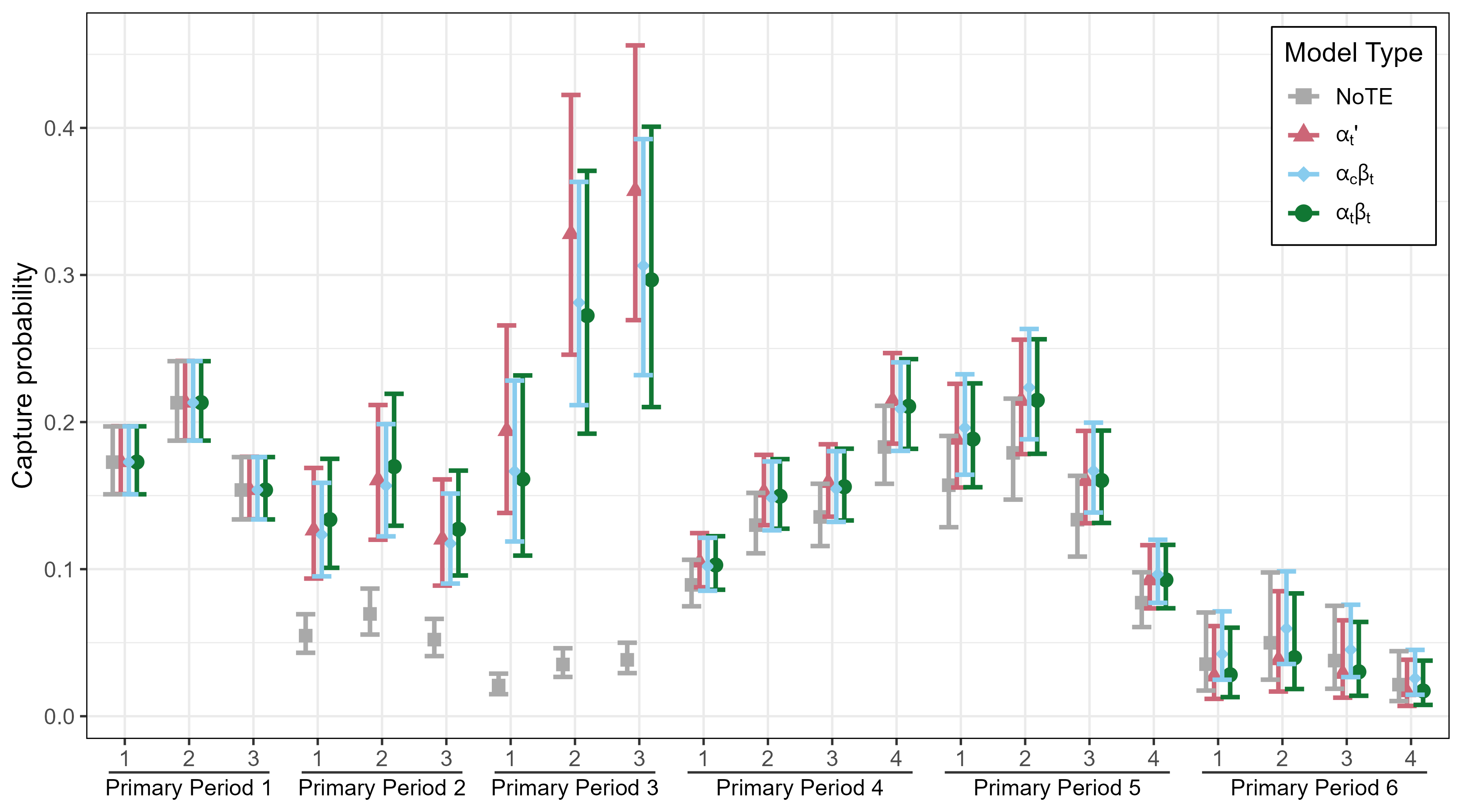}
    \caption{Estimation results of capture probabilities for the mantella data from the NoTE model and TE models $\alpha_t', \alpha_c\beta_t,$ and $\alpha_t\beta_t$.}
    \label{fig:p}
\end{figure}

Figure \ref{fig:p} shows a comparison of the estimates of capture probabilities between the NoTE model and three TE models $\alpha_t\beta_t, \alpha_c\beta_t,$ and $\alpha_t'$ with the lowest AIC values. We first note that the estimates of capture probabilities for the first primary period are exactly the same for all models with or without TE. This is because individuals need to join the super-population first before becoming temporary emigrants in subsequent periods, and thus TE does not occur before the first primary period and does not affect the estimation of capture probabilities in the first period. A direct consequence of accounting for temporary emigration is that TE models produce higher capture probabilities compared to the NoTE model, although the differences are minor in the final primary period. This occurs because when TE is not allowed, individuals that have emigrated are assumed to be present in the sampling area but undetected, leading to the underestimation of capture probabilities. Higher capture probability estimates in the TE models result in lower abundance estimates as shown in Table \ref{table:aics}. When individuals have a higher probability of being detected, fewer individuals are undetected, which in turn reduces the estimated abundance.\\

We note that differences in capture probabilities between NoTE and TE models are largest in primary period 3. As shown in Table \ref{tab:alphabeta}, the estimated probability of temporary emigration before this period is high ($\hat{\alpha_2}=0.85$), while the probability of return is low ($\hat{\beta_2}=0.01$). This results in many individuals being in state 3 during period 3, inflating the estimated capture probabilities relative to the NoTE model. In later periods, higher return probabilities reduce the number of individuals in state 3, leading to smaller differences between models. Additionally, we note that the 95\% confidence intervals for capture probabilities are generally wider for the TE models, as these models introduce additional complexity. Even after accounting for temporary emigration, the overall observed pattern in the capture probability estimates remains somewhat periodic, and there is noticeable variation in capture probabilities within primary periods $3-5$. These observations might suggest a potential violation of the assumption of population closure within the primary periods. Some research is ongoing to relax the population closure assumption, but this is beyond the scope of this manuscript.  

%% file: discussion.tex
\FloatBarrier
\section{Discussion}\label{dis}
This paper presents a flexible latent multinomial modelling framework for CR data with temporary emigration. Compared to existing TE models in the literature, the proposed method accounts for temporary emigration while relaxing the assumption of perfect individual identification, which is often not met in modern CR studies. As seen from the simulation studies, all model parameters except for those that are confounded can be estimated reliably. The mantella data application shows that accounting for temporary emigration results in a noticeable improvement in model fit. This highlights the benefits of the new model for handling temporary emigration in CR data with latent identification, and one is encouraged to consider TE models if there is any evidence of temporary emigration. \\

Reflecting on the structure of the motivating mantella dataset, the modelling framework was developed for batch-marking CR data collected using the robust design. The advantage of robust design is that it offers additional information through secondary sampling occasions, which helps to inform parameter estimation. Unfortunately, many datasets are not collected under robust design. As outlined in Section \ref{modelvariants}, the presented framework is applicable to other study designs, such as closed populations, open populations, and open robust design. The closed-population scenario is a special case of the robust design, thus there will not be any issues applying the framework shown in this paper. For open-population models such as the Cormack-Jolly-Seber model, Markovian emigration can be handled using the multi-state model of \cite{fujiwara2002general}. As noted by \cite{schaub2004estimating} it may require additional constraints on some model parameters, but only when survival, recruitment, and emigration parameters are all time-dependent. Some preliminary results under the open robust design similar to \citet{schwarz1997estimating} suggest that the model could perform well without requiring too many constraints; however, further investigation is needed to explore this, which is beyond the scope of this paper. Note that batch marking is just one example of latent identification in CR studies, and as shown in Section \ref{modelvariants}, the framework applies to other scenarios such as misidentification. Following the same logic, it could also work for CR data integration problems \citep[e.g.,][]{bonner2013mark,mcclintock2013integrated}. \\

Another advantage of the proposed framework is that it is not limited to specific emigration structures, and as we have shown, both the completely random and Markovian emigration mechanics fit naturally into the framework. The framework would also be suitable for other types of emigration - for example, one might assume trap-dependent temporary emigration \citep{kendall1997estimating}, where an individual's emigration probability on one occasion depends on whether it was captured on the previous occasion.\\

An implicit assumption underlying the proposed model is that survival probability does not depend on an individual’s emigration status (i.e. in or out of the population). As discussed in \citet{kendall1997estimating}, it is possible that there are some external factors affecting the survival of an individual while it stays out of the study area. This could easily be accounted for in our model by specifying different survival probabilities according to if the individual is in or out of the population. However, as noted by \cite{kendall1997estimating}, further investigation is needed to see whether survival probabilities for individuals out of the study area could be estimated.

%% file: acknowledgement.tex
\section*{Acknowledgements}
This work was funded by the Carnegie Trust for the Universities of Scotland. We would like to thank Madagasikara Voakajy for providing the mantella data. We are grateful to Dr. Simon Bonner and Prof. Rachel McCrea for their helpful discussions on the initial development of the work. Thanks to Dr. Simon Bonner and Dr. Craig Wilkie for their helpful comments on earlier versions of this manuscript.

%% file: supplementary_material/sm_main.tex
% \documentclass[a4paper, 12pt]{article}
% \usepackage{comment} 
% \usepackage{float}
% \usepackage{graphicx}
% \usepackage{caption}
% \usepackage{color,soul}
% \usepackage{fullpage} 
% \usepackage{natbib}
% \usepackage[T1]{fontenc}
% \usepackage{upquote}
% \usepackage[dvipsnames]{xcolor}
% \bibliographystyle{abbrvnat}
% \setcitestyle{authoryear,open={(},close={)}}
% \usepackage{hyperref}
% \usepackage{amsmath,amsfonts,amssymb}
% \usepackage{longtable}
% \setlength{\parindent}{0pt}
% \usepackage{color,soul}
% \usepackage[toc,page]{appendix}
% \usepackage{booktabs}
% \usepackage{nicematrix}
% \usepackage[section]{placeins}
% \usepackage{multirow}
% \usepackage{tikz}
% \usetikzlibrary{fit}
% \usepackage{colortbl}

\renewcommand{\thesubsection}{\Alph{section}.\arabic{subsection}}

% highlighter colours
\DeclareRobustCommand{\hlcyan}[1]{{\sethlcolor{SkyBlue}\hl{#1}}}
\DeclareRobustCommand{\lime}[1]{{\sethlcolor{lime}\hl{#1}}}
\DeclareRobustCommand{\gold}[1]{{\sethlcolor{Goldenrod}\hl{#1}}}
\DeclareRobustCommand{\apricot}[1]{{\sethlcolor{Apricot}\hl{#1}}}
\DeclareRobustCommand{\orch}[1]{{\sethlcolor{Thistle}\hl{#1}}}

% % 1-inch margins all around
% \usepackage[margin=1in]{geometry}
% \doublespacing

% \begin{document}
\title{\Large Supplementary Materials for ``Estimating temporary emigration from capture-recapture data in the presence of latent identification'' by Katarina Skopalova, Jafet Belmont, and Wei Zhang}

\date{}
\maketitle
%\tableofcontents
\appendix

\section*{Web Appendix A: Additional Simulation Results}

\addcontentsline{toc}{section}{Web Appendix A: Additional Simulation Results}
\setcounter{section}{1} 
This web appendix presents simulation results for five temporary emigration (TE) models: the completely random emigration model ($\alpha_t'$), and four Markovian models ($\alpha_t\beta_t$, $\alpha_c\beta_c$, $\alpha_c\beta_t$, and $\alpha_t\beta_c$). Each model was fitted to data with individual identification (ID) as well as to batch-mark data (BM). We provide an overview of the main findings below, while detailed comparison graphs for each model are given in the corresponding subsections. For completeness, the BM results for model $\alpha_t\beta_t$ originally shown in the main paper are also reproduced here. The parameter settings were: $N=5000$, $p_{kl}=0.4$, $\phi_k=0.9$, $\gamma_k=1/6$, $\alpha_k'=0.2$, $\alpha_k=0.2$, and $\beta_k=0.7$. We simulated 100 datasets under the robust design with $K=6$ primary periods and $T_k=2$ secondary occasions per period. Results for capture and entry probabilities are omitted, as these were consistently estimated with high accuracy and precision across all models.\\

Under the completely random emigration model, both survival and emigration parameters are estimated without bias in both scenarios, with good 95\% CI coverage. Confidence intervals are relatively narrow in each case, though mean CI widths are about twice as narrow with individual identification (ID) compared to batch-marking (BM). For the Markovian emigration models, parameter estimates are generally more precise under ID than under BM. When both emigration and immigration parameters are held constant, ID and BM yield similarly accurate estimates that closely match the true values, with the exception of $\beta$: under ID, $\beta$ is estimated without bias, whereas under BM it tends to be slightly overestimated. Overall, the $\alpha_c\beta_c$ model produces more accurate and precise estimates than the time-dependent Markovian models, as constraining the parameters reduces model complexity.\\

In the remaining Markovian models (where either $\alpha$, $\beta$, or both are time-dependent), ID generally yields higher precision. In terms of bias, ID produces more accurate estimates for $\alpha$ and $\beta$, while estimates for $\phi$ and $N$ are typically comparable across scenarios. In the fully time-dependent model, mean survival estimates are similar under ID and BM for earlier primary periods, with noticeable bias only emerging in later primary periods, particularly for $\phi_4$. Even then, survival is only slightly underestimated under both scenarios and remains close to the true value. Emigration estimates also closely match the true values, though $\alpha_4$ is slightly underestimated in both cases (less so under ID). For $\beta$, mean estimates under ID show slight underestimation with relatively wide CIs. As discussed in the main paper, this is a consequence of the low emigration probability. Increasing $\alpha$ mitigates this issue, as illustrated by fitting the $\alpha_t\beta_t$ model with $\alpha=0.4$ (see Figures \ref{fig:atbt_n_high} and \ref{fig:atbt_alpha_beta_phi_high} in Subsection \ref{atbt}).

\subsection{Completely random emigration model $\alpha_t'$}
\input{CR}

\subsection{Markovian emigration model $\alpha_t\beta_t$}
\input{atbt}\label{atbt}

\subsection{Markovian emigration model $\alpha_c\beta_c$}
\input{acbc}

\subsection{Markovian emigration model $\alpha_c\beta_t$}

\input{acbt}

\subsection{Markovian emigration model $\alpha_t\beta_c$}

\input{atbc}

% \end{document}

%% file: supplementary_material/CR.tex
\begin{figure}[H]
    \centering
    \includegraphics[width=1\linewidth]{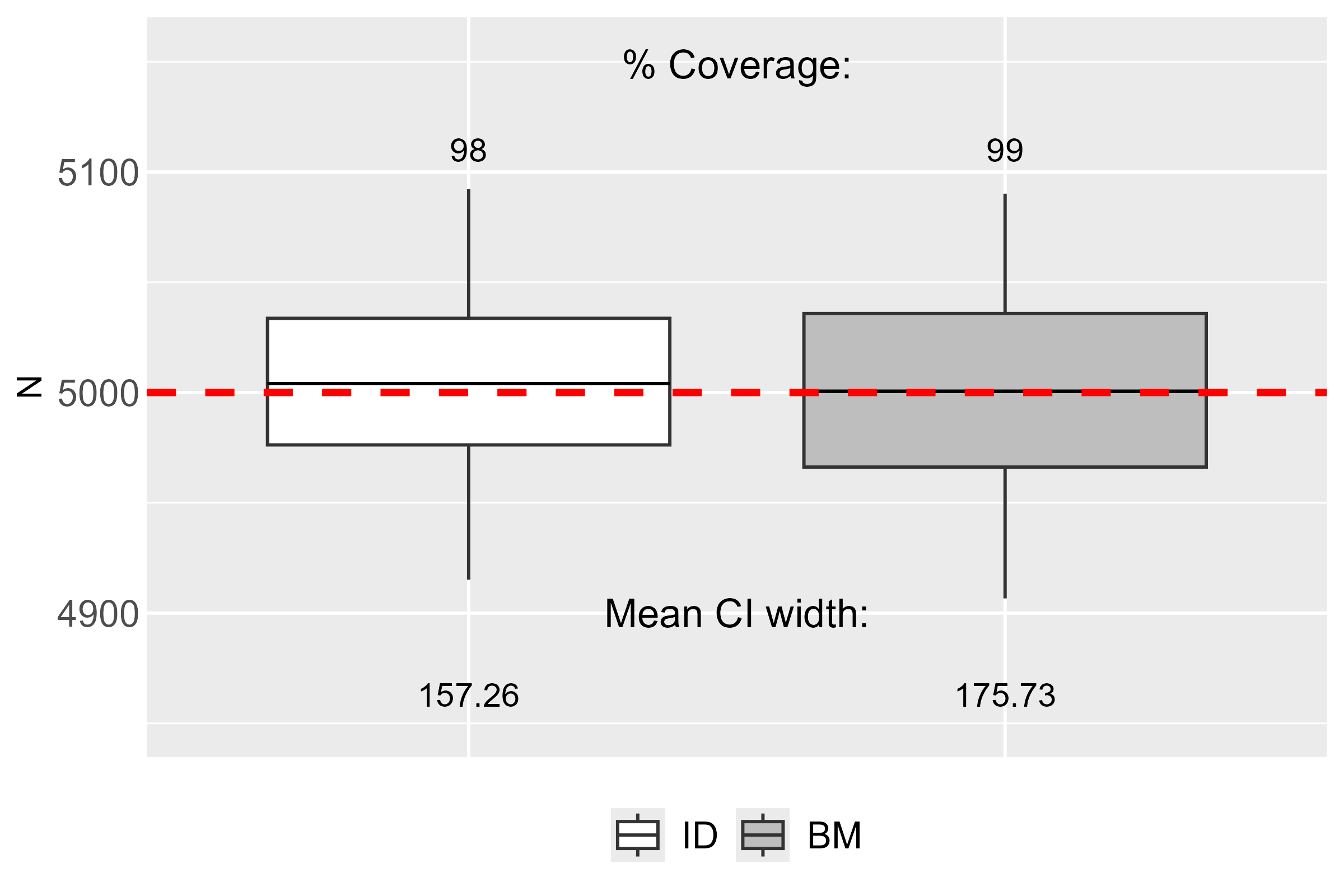}
    \caption{Boxplots of maximum likelihood estimates of the super-population size $N$ obtained by fitting the completely random emigration model $\alpha_t'$ to 100 ID and BM datasets. The horizontal line inside each box represents the mean estimate across 100 simulations, while the gray dashed line indicates the true parameter value. 95\% CI coverage rates and mean CI widths are shown above and below each boxplot, respectively.}
    \label{fig:cr_n}
\end{figure}

\begin{figure}[H]
    \centering
    \includegraphics[width=1\linewidth]{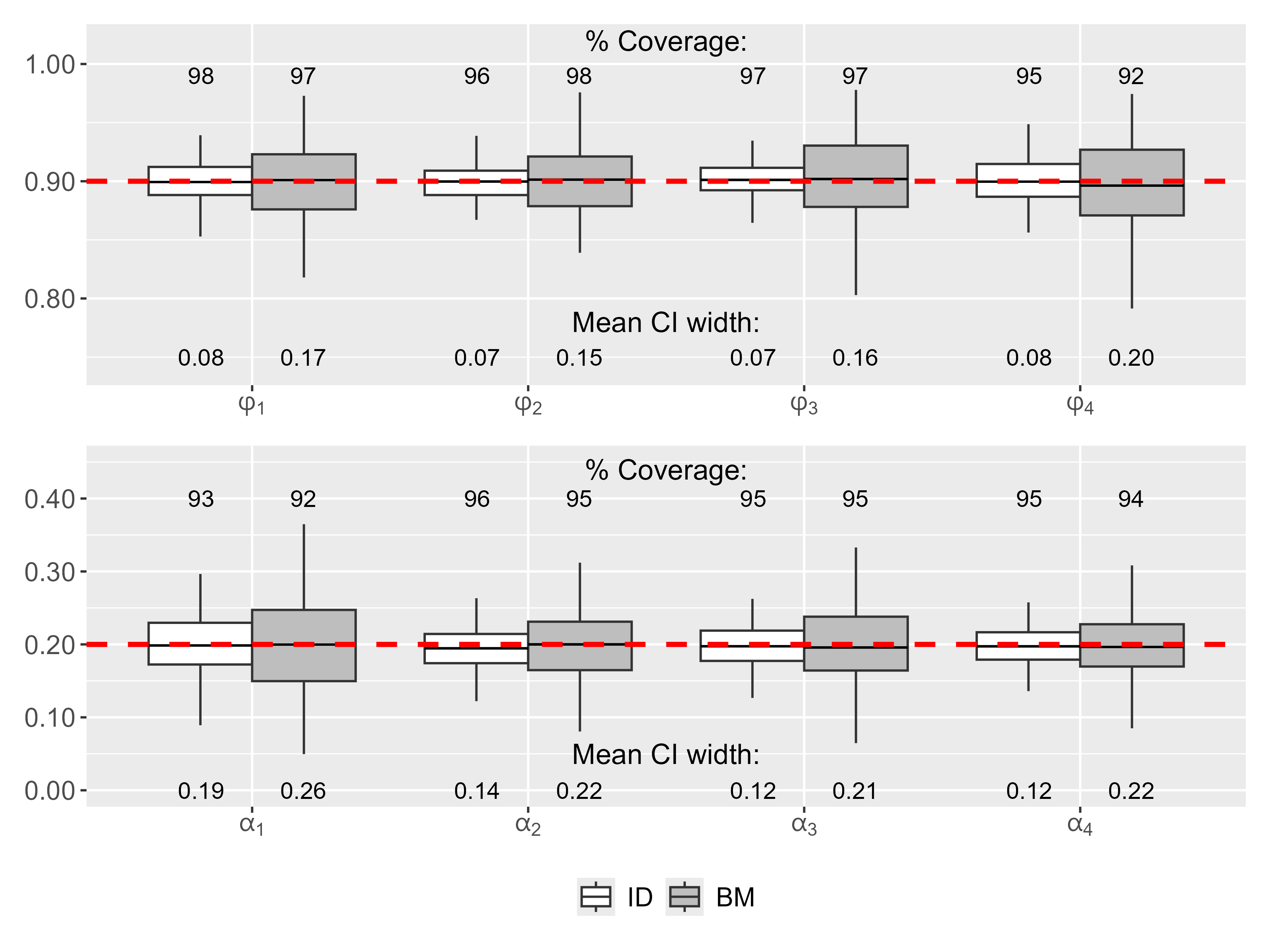}
    \caption{Boxplots of maximum likelihood estimates of survival probabilities $\boldsymbol{\phi}$ (top) and emigration probabilities $\boldsymbol{\alpha'}$ (bottom) obtained by fitting the completely random emigration model $\alpha_t'$ to 100 ID and BM datasets. The horizontal line inside each box represents the mean estimate across 100 simulations, while the gray dashed line indicates the true parameter value. 95\% CI coverage rates and mean CI widths are shown above and below each boxplot, respectively.}
    \label{fig:cr_phi_alpha}
\end{figure}

%% file: supplementary_material/atbt.tex
\begin{figure}[H]
    \centering
    \includegraphics[width=1\linewidth]{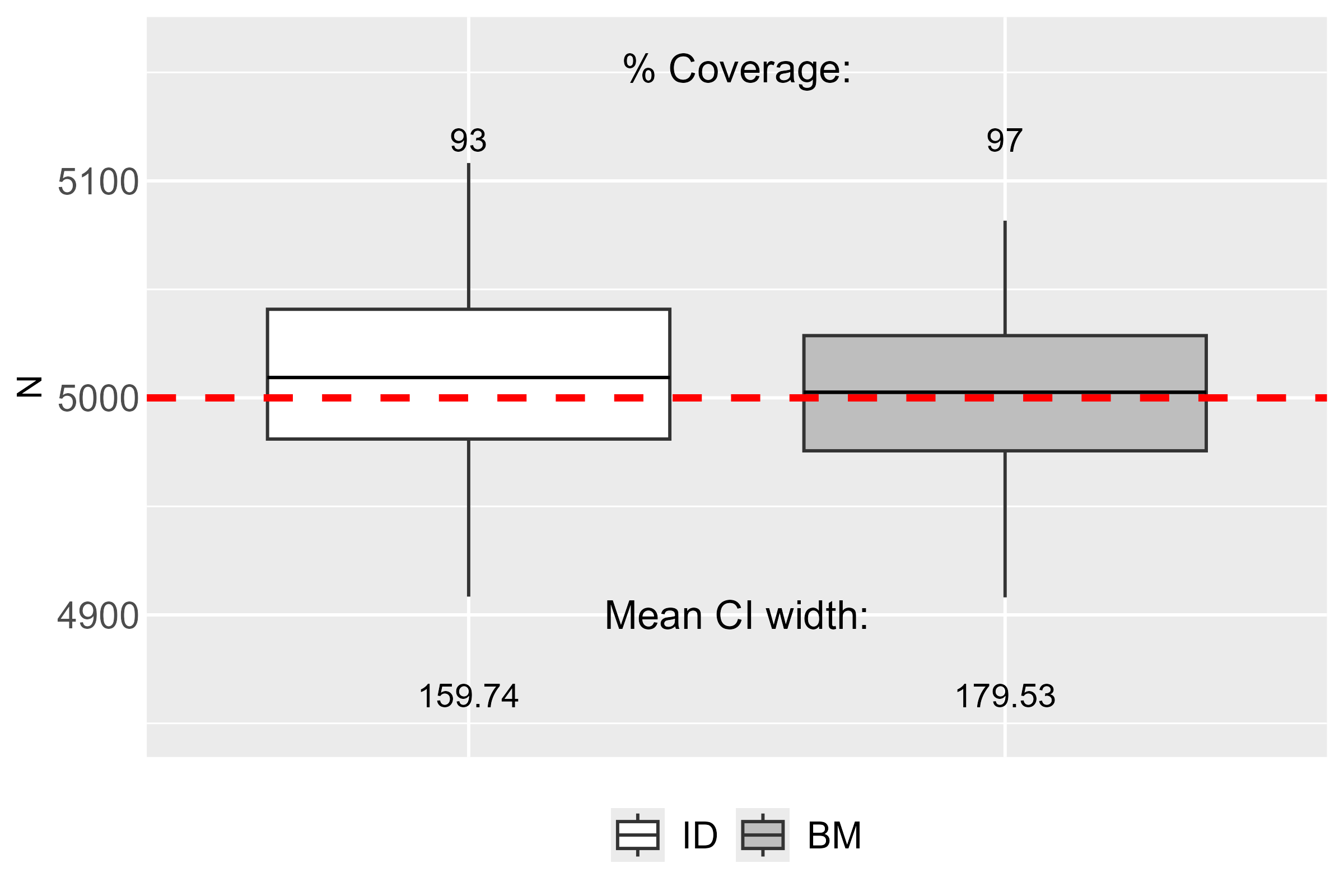}
    \caption{Boxplots of maximum likelihood estimates of the super-population size $N$ obtained by fitting the fully time-dependent Markovian model $\alpha_t\beta_t$ to 100 ID and BM datasets, with $\alpha_k=0.2$. The horizontal line inside each box represents the mean estimate across 100 simulations, while the gray dashed line indicates the true parameter value. 95\% CI coverage rates and mean CI widths are shown above and below each boxplot, respectively.}
    \label{fig:atbt_n_low}
\end{figure}

\begin{figure}[H]
    \centering
    \includegraphics[width=1\linewidth]{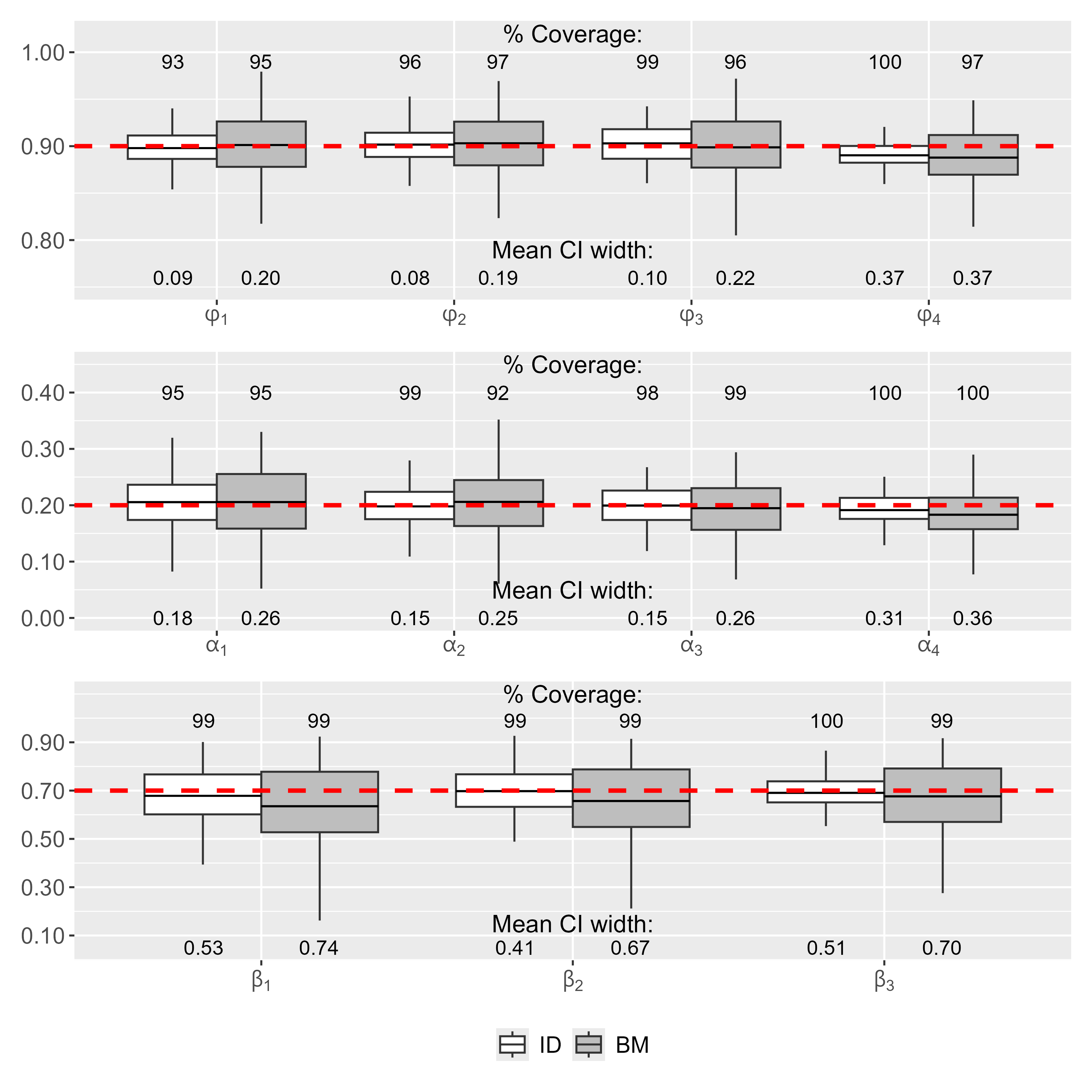}
    \caption{Boxplots of maximum likelihood estimates of survival probabilities $\boldsymbol{\phi}$ (top), emigration probabilities $\boldsymbol{\alpha}$ (middle), and immigration probabilities $\boldsymbol{\beta}$ (bottom) obtained by fitting the fully time-dependent Markovian model $\alpha_t\beta_t$ to 100 ID and BM datasets, with $\alpha_k=0.2$. The horizontal line inside each box represents the mean estimate across 100 simulations, while the gray dashed line indicates the true parameter value. 95\% CI coverage rates and mean CI widths are shown above and below each boxplot, respectively.}
    \label{fig:atbt_vsetko_low}
\end{figure}

\begin{figure}[H]
    \centering
    \includegraphics[width=1\linewidth]{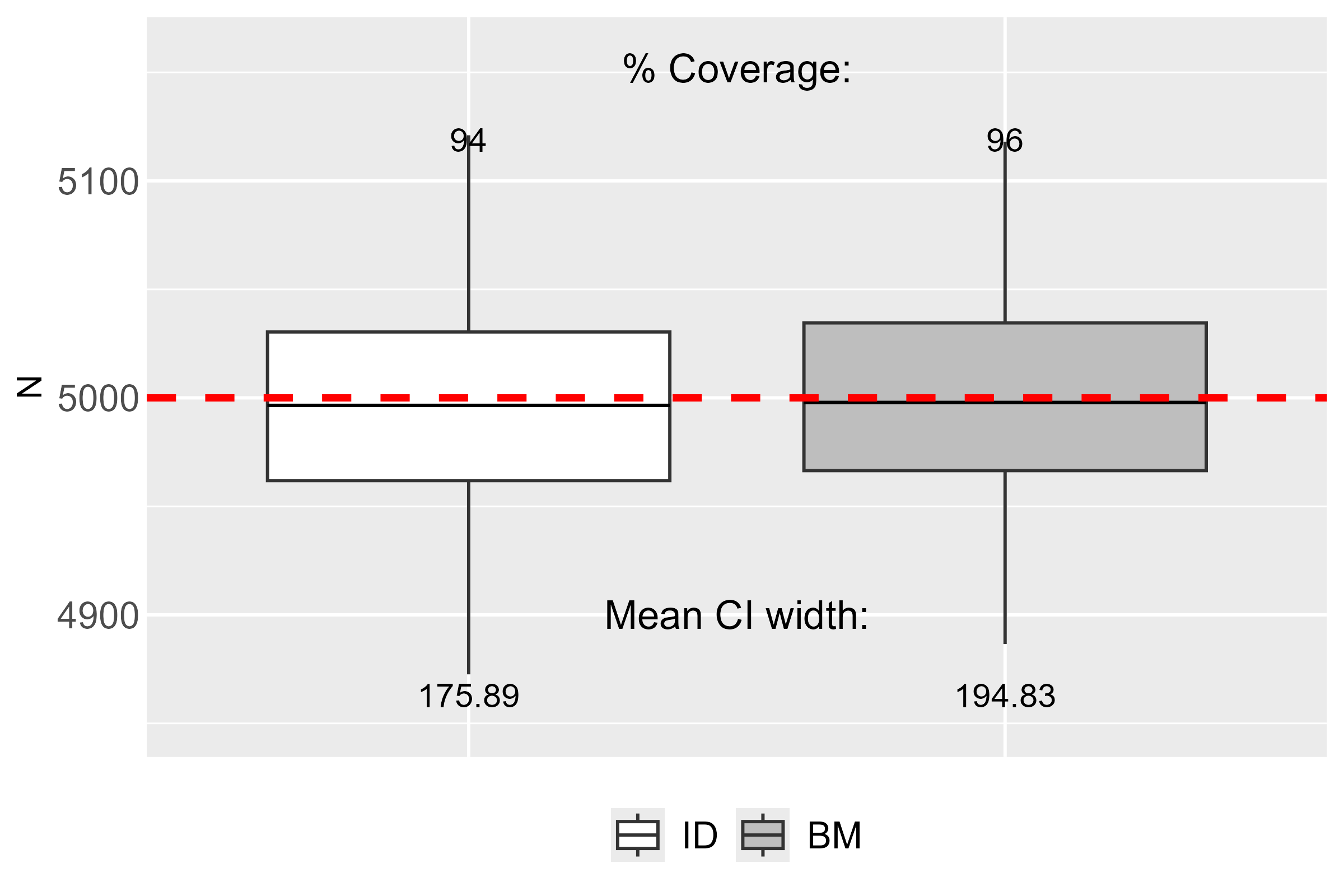}
    \caption{Boxplots of maximum likelihood estimates of the super-population size $N$ obtained by fitting the fully time-dependent Markovian model $\alpha_t\beta_t$ to 100 ID and BM datasets, with $\alpha_k=0.4$. The horizontal line inside each box represents the mean estimate across 100 simulations, while the gray dashed line indicates the true parameter value. 95\% CI coverage rates and mean CI widths are shown above and below each boxplot, respectively.}
    \label{fig:atbt_n_high}
\end{figure}

\begin{figure}[H]
    \centering
    \includegraphics[width=1\linewidth]{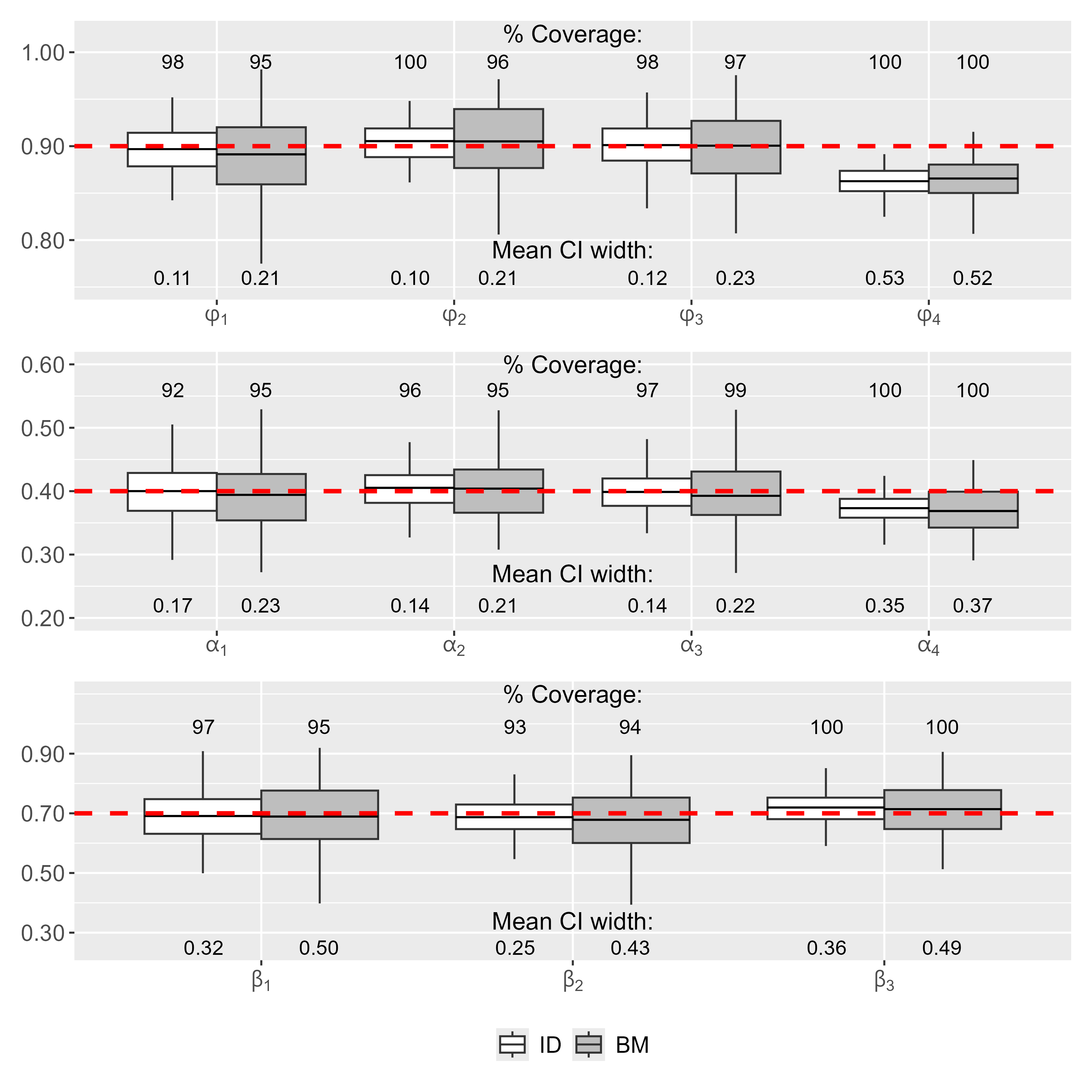}
    \caption{Boxplots of maximum likelihood estimates of survival probabilities $\boldsymbol{\phi}$ (top), emigration probabilities $\boldsymbol{\alpha}$ (middle), and immigration probabilities $\boldsymbol{\beta}$ (bottom) obtained by fitting the fully time-dependent Markovian model $\alpha_t\beta_t$ to 100 ID and BM datasets, with $\alpha_k=0.4$. The horizontal line inside each box represents the mean estimate across 100 simulations, while the gray dashed line indicates the true parameter value. 95\% CI coverage rates and mean CI widths are shown above and below each boxplot, respectively.}
    \label{fig:atbt_alpha_beta_phi_high}
\end{figure}

%% file: supplementary_material/acbc.tex
\begin{figure}[H]
    \centering
    \includegraphics[width=1\linewidth]{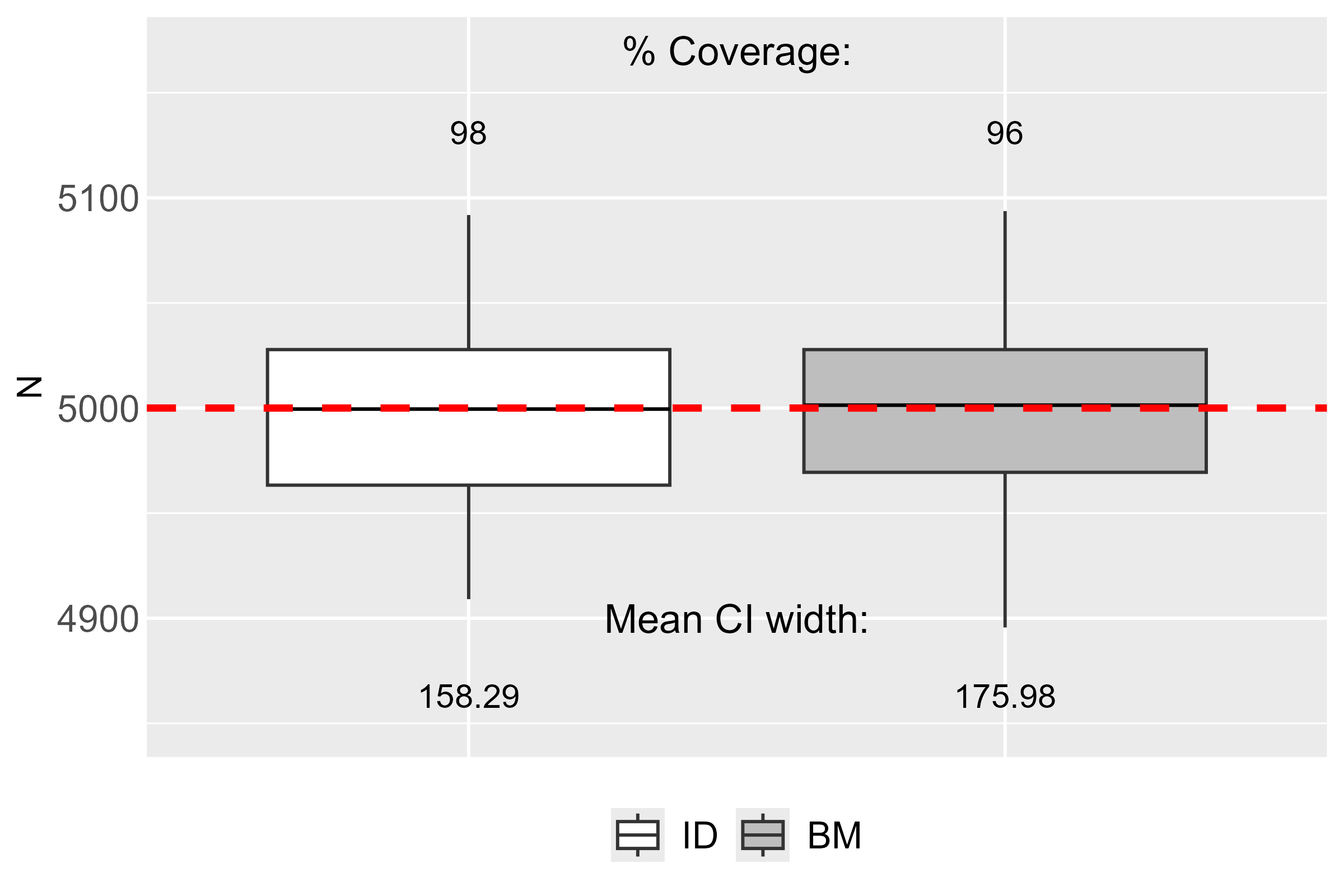}
    \caption{Boxplots of maximum likelihood estimates of the super-population size $N$ obtained by fitting the Markovian emigration model $\alpha_c\beta_c$ to 100 ID and BM datasets. The horizontal line inside each box represents the mean estimate across 100 simulations, while the gray dashed line indicates the true parameter value. 95\% CI coverage rates and mean CI widths are shown above and below each boxplot, respectively.}
    \label{fig:acbc_n}
\end{figure}

\begin{figure}[H]
    \centering
    \includegraphics[width=1\linewidth]{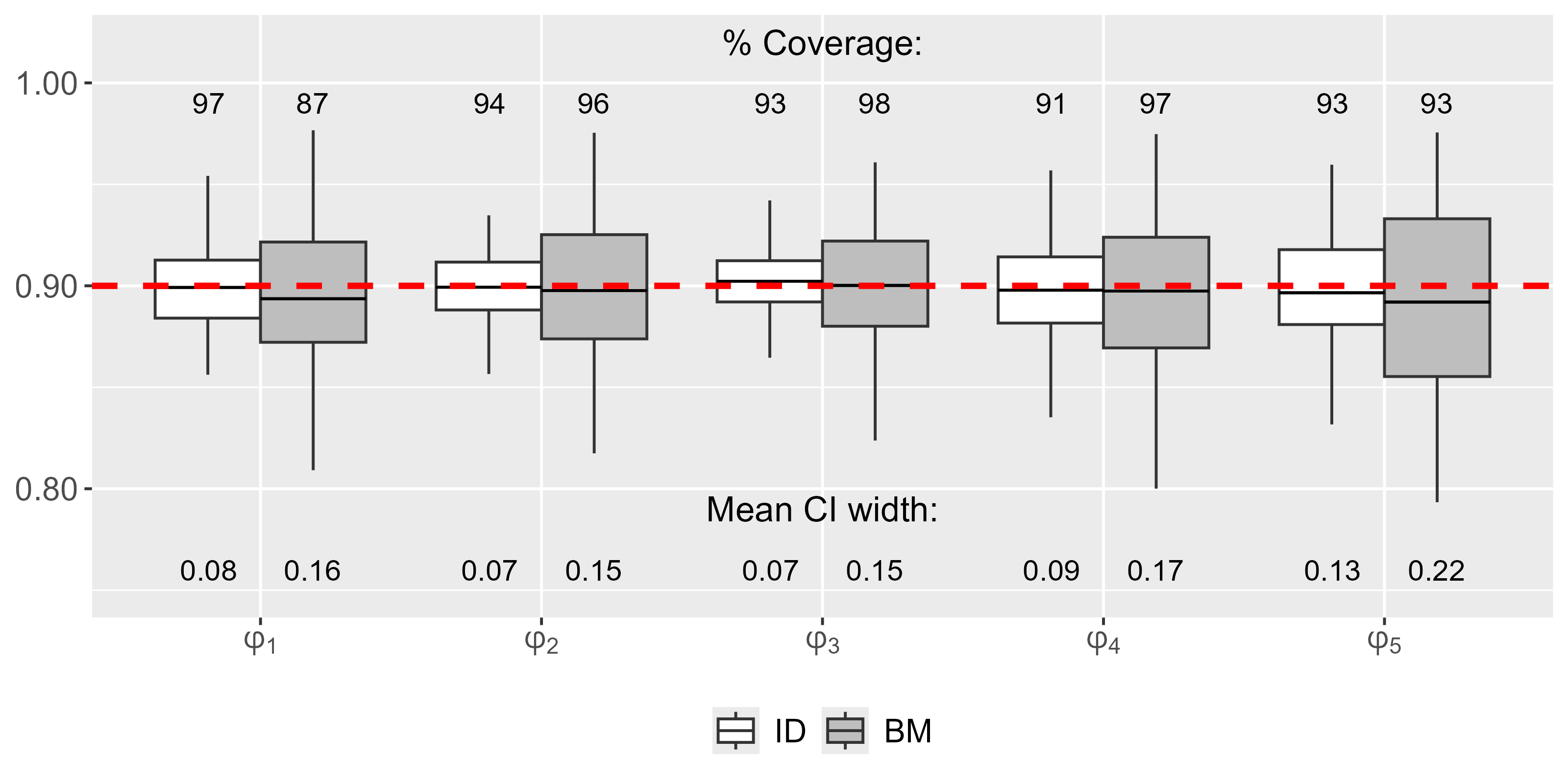}
    \caption{Boxplots of maximum likelihood estimates of survival probabilities $\boldsymbol{\phi}$ obtained by fitting the Markovian emigration model $\alpha_c\beta_c$ to 100 ID and BM datasets. The horizontal line inside each box represents the mean estimate across 100 simulations, while the gray dashed line indicates the true parameter value. 95\% CI coverage rates and mean CI widths are shown above and below each boxplot, respectively.}
    \label{fig:acbc_phi}
\end{figure}

\begin{figure}[H]
    \centering
    \includegraphics[width=1\linewidth]{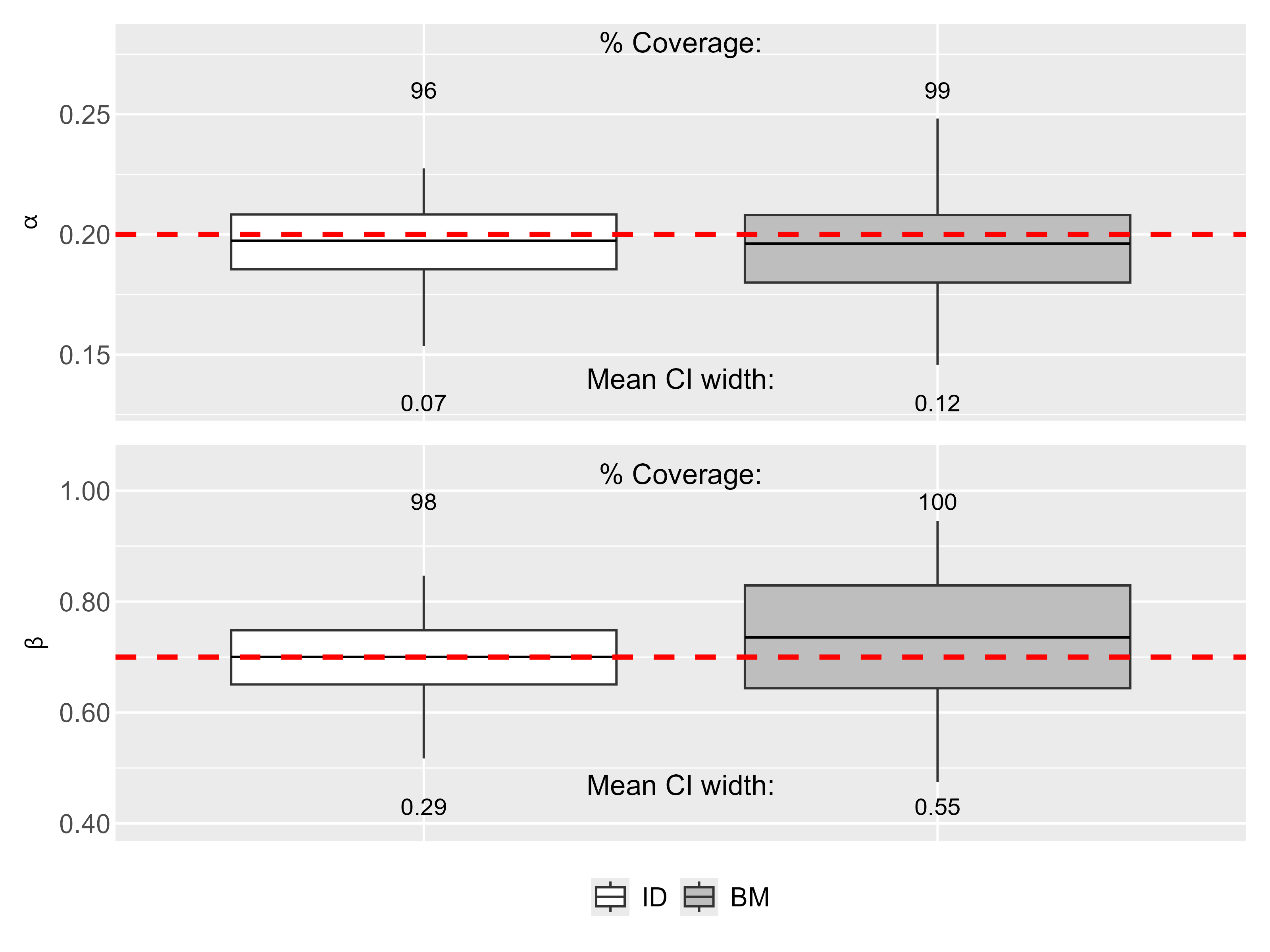}
    \caption{Boxplots of maximum likelihood estimates of emigration probability $\alpha$ (top) and immigration probability $\beta$ (bottom) obtained by fitting the Markovian emigration model $\alpha_c\beta_c$ to 100 ID and BM datasets. The horizontal line inside each box represents the mean estimate across 100 simulations, while the gray dashed line indicates the true parameter value. 95\% CI coverage rates and mean CI widths are shown above and below each boxplot, respectively.}
    \label{fig:acbc_ab}
\end{figure}

%% file: supplementary_material/acbt.tex
\begin{figure}[H]
    \centering
    \includegraphics[width=1\linewidth]{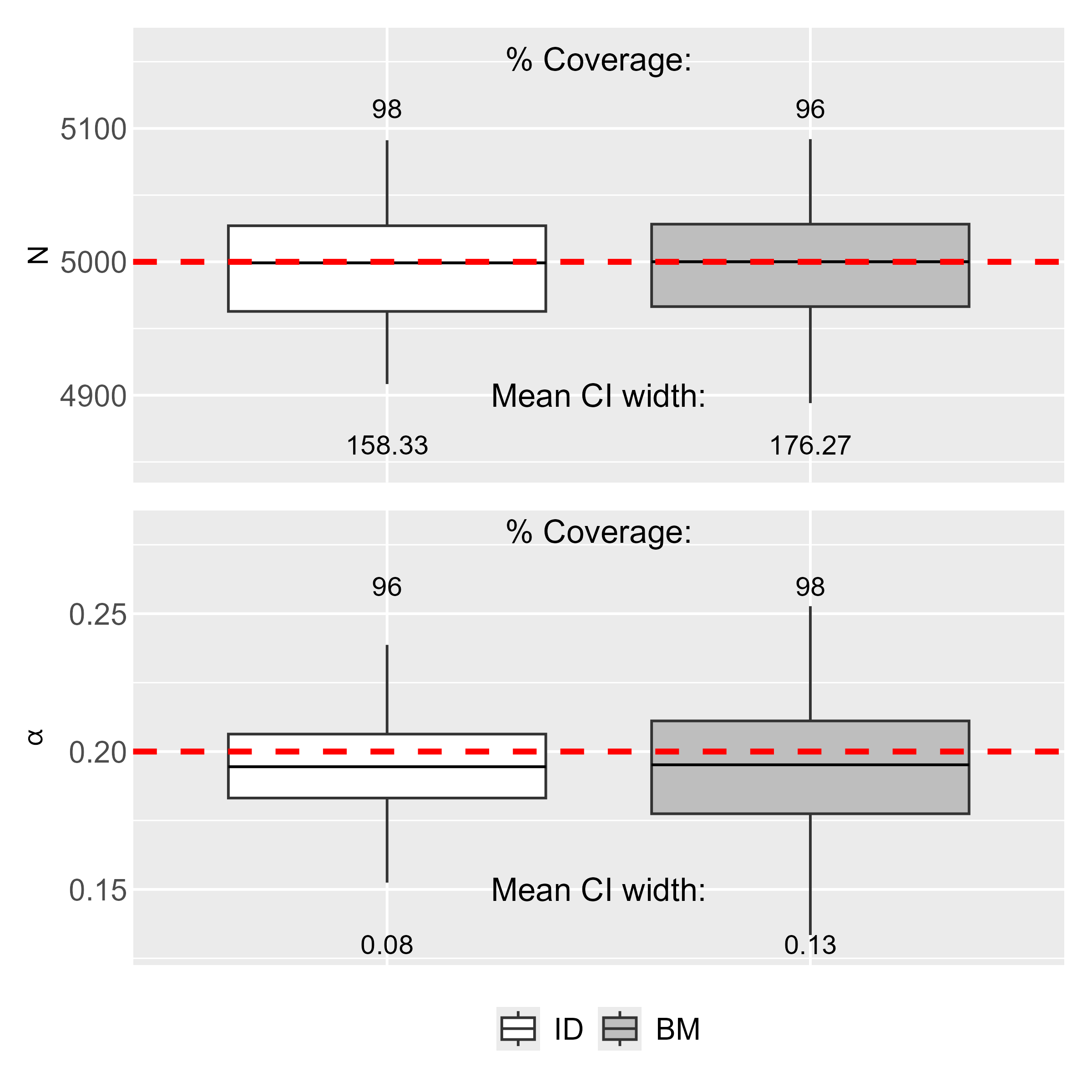}
    \caption{Boxplots of maximum likelihood estimates of the super-population size $N$ and emigration probability $\alpha$ obtained by fitting the Markovian emigration model $\alpha_c\beta_t$ to 100 ID and BM datasets. The horizontal line inside each box represents the mean estimate across 100 simulations, while the gray dashed line indicates the true parameter value. 95\% CI coverage rates and mean CI widths are shown above and below each boxplot, respectively.}
    \label{fig:acbt_n_alpha}
\end{figure}

\begin{figure}[H]
    \centering
    \includegraphics[width=1\linewidth]{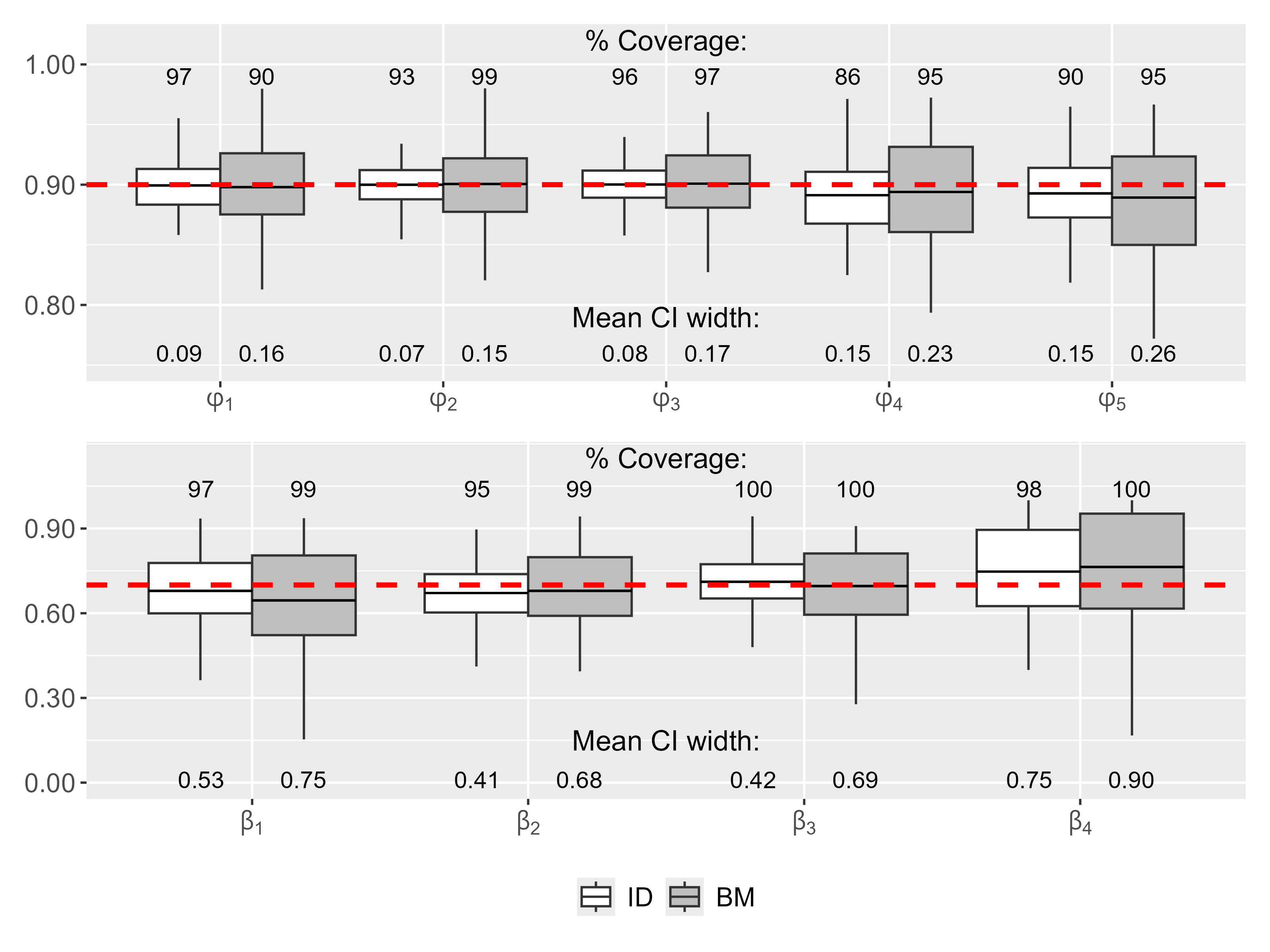}
    \caption{Boxplots of maximum likelihood estimates of survival probabilities $\boldsymbol{\phi}$ and immigration probabilities $\boldsymbol{\beta}$ obtained by fitting the Markovian emigration model $\alpha_c\beta_t$ to 100 ID and BM datasets. The horizontal line inside each box represents the mean estimate across 100 simulations, while the gray dashed line indicates the true parameter value. 95\% CI coverage rates and mean CI widths are shown above and below each boxplot, respectively.}
    \label{fig:acbt_phi_beta}
\end{figure}

%% file: supplementary_material/atbc.tex
\begin{figure}[H]
    \centering
    \includegraphics[width=1\linewidth]{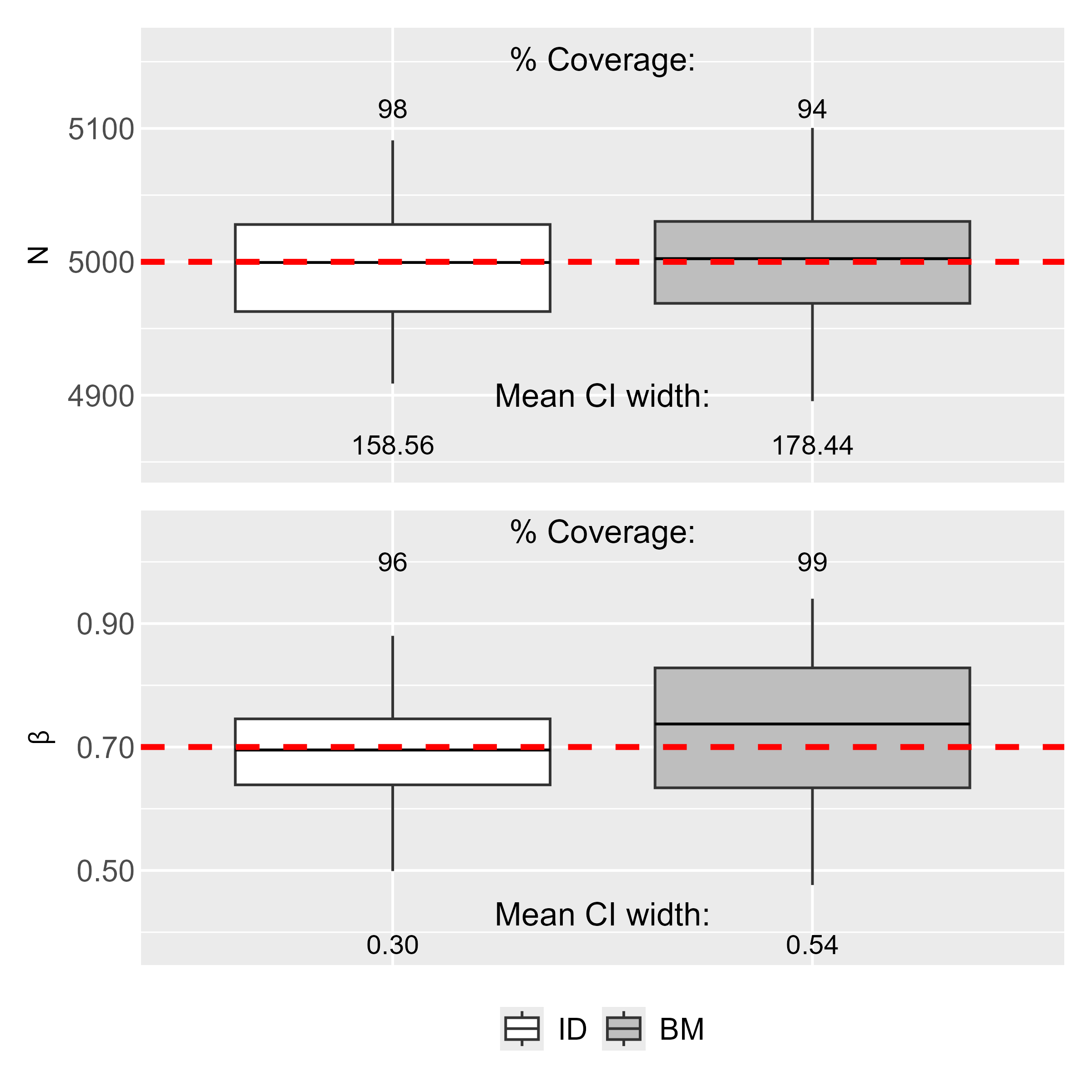}
    \caption{Boxplots of maximum likelihood estimates of the super-population size $N$ and immigration probability $\beta$ obtained by fitting the Markovian emigration model $\alpha_t\beta_c$ to 100 ID and BM datasets. The horizontal line inside each box represents the mean estimate across 100 simulations, while the gray dashed line indicates the true parameter value. 95\% CI coverage rates and mean CI widths are shown above and below each boxplot, respectively.}
    \label{fig:atbc_n_beta}
\end{figure}

\begin{figure}[H]
    \centering
    \includegraphics[width=1\linewidth]{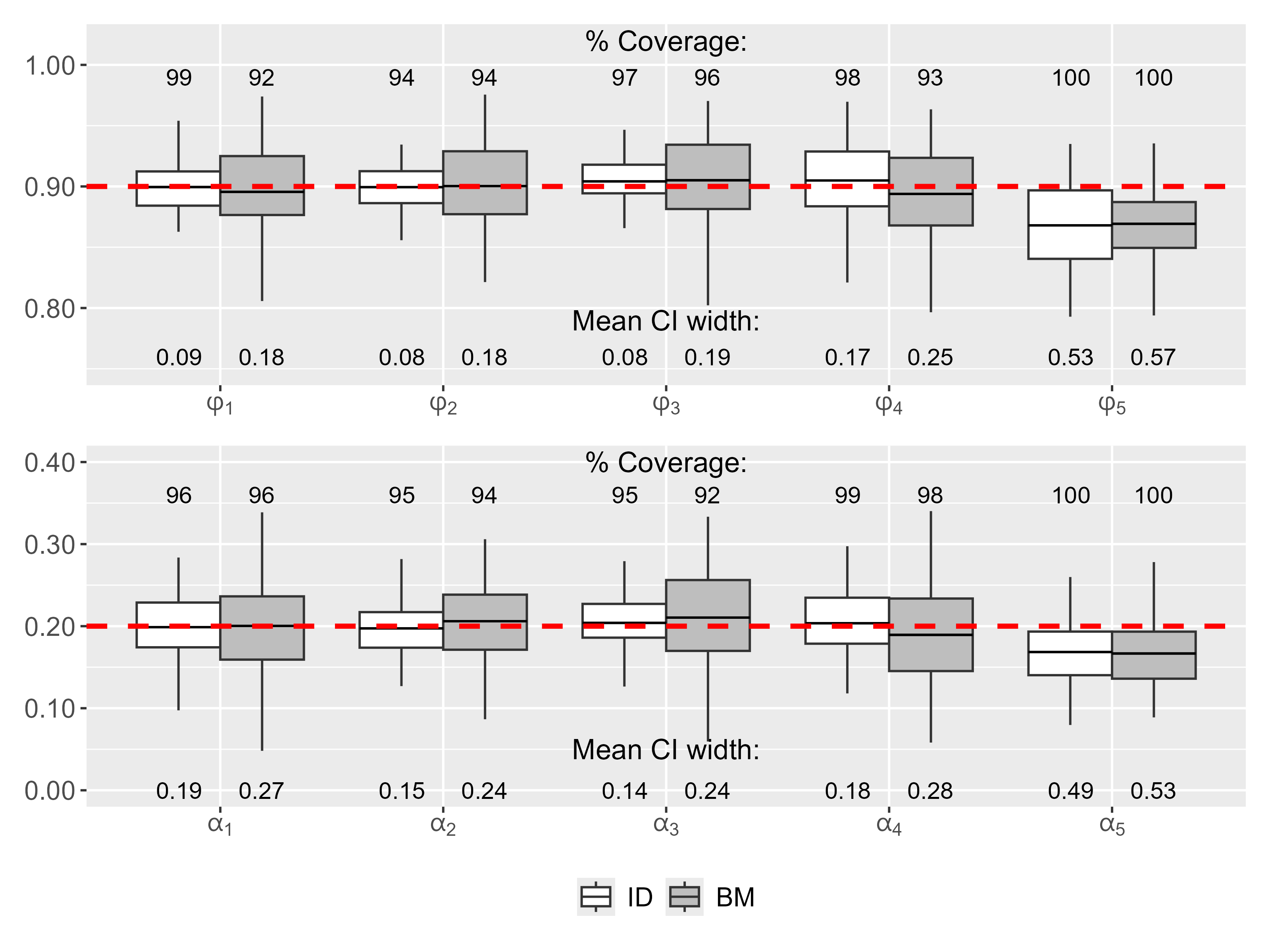}
    \caption{Boxplots of maximum likelihood estimates of survival probabilities $\boldsymbol{\phi}$ and emigration probabilities $\boldsymbol{\alpha}$ obtained by fitting the Markovian emigration model $\alpha_t\beta_c$ to 100 ID and BM datasets. The horizontal line inside each box represents the mean estimate across 100 simulations, while the gray dashed line indicates the true parameter value. 95\% CI coverage rates and mean CI widths are shown above and below each boxplot, respectively.}
    \label{fig:atbc_phi_alpha}
\end{figure}